\documentclass[final,3p,times,twocolumn,authoryear]{elsarticle}

\usepackage[utf8]{inputenc}
\usepackage{amssymb}
\usepackage{xcolor}
\usepackage{graphicx}	
\usepackage{amsmath}	
\usepackage{hyperref}
\usepackage{multirow}
\usepackage{wasysym}
\usepackage{soul}
\usepackage{newtxtext,newtxmath}

\begin{document}

\newcommand{\gpmDM}{\ensuremath{273.5}}
\newcommand{\DMerror}{\ensuremath{2.6}}
\newcommand{\DMunit}{pc\,cm\ensuremath{^{-3}}}
\newcommand{\gpmDist}{\ensuremath{5.7}}
\newcommand{\Disterror}{\ensuremath{2.9}}
\newcommand{\Distunit}{kpc}
\newcommand{\gpmRM}{\ensuremath{-573}}
\newcommand{\RMerror}{\ensuremath{1}}
\newcommand{\RMunit}{rad\,m\ensuremath{^{-2}}}
\newcommand{\gpmPlong}{\ensuremath{1318.1957}}
\newcommand{\gpmPerr}{\ensuremath{0.0002}}
\newcommand{\gpmP}{\ensuremath{1318}}
\newcommand{\gpmFlong}{\ensuremath{0.000758612(7)}}
\newcommand{\gpmFerr}{\ensuremath{1.2\times10^{-10}}}
\newcommand{\gpmFdotlong}{\ensuremath{5\times10^{-20}}}

\newcommand{\gpmalpha}{\ensuremath{-3.17}}
\newcommand{\gpmalphaerr}{\ensuremath{0.06}}
\newcommand{\gpmq}{\ensuremath{-0.56}}
\newcommand{\gpmqerr}{\ensuremath{0.03}}

\newcommand{\ndetections}{71}
\newcommand{\gpmPdot}{\ensuremath{3.6\times10^{-13}}}
\newcommand{\gpmPdotts}{\ensuremath{9.5\times10^{-13}}}
\newcommand{\Pdotunit}{\,s\,s\ensuremath{^{-1}}}

\newcommand{\ergpers}{erg\,s\ensuremath{^{-1}}}
\newcommand{\flux}{erg\,cm\ensuremath{^{-2}}\,s\ensuremath{^{-1}}}

\begin{frontmatter}



\title{Long Period Transients (LPTs): a comprehensive review}


\author{Nanda Rea$^{a,b}$, Natasha Hurley-Walker$^{c}$ \& Manisha Caleb$^{d,e}$ } 

\affiliation[label1]{organization={Institute of Space Sciences (ICE), CSIC},
            addressline={Campus UAB, Carrer de Can Magrans} s/n, 
            city={Bellaterra (Barcelona)},
            postcode={E-08193}, 
            country={Spain}}

\affiliation[label2]{organization={Institut d’Estudis Espacials de Catalunya (IEEC)},
            city={Castelldefels (Barcelona)},
            postcode={08860}, 
            country={Spain}}

\affiliation[label3]{organization={International Centre for Radio Astronomy Research, Curtin University},
            addressline={Kent Street}, 
            city={Bentley WA},
            postcode={6102}, 
            country={Australia}}

\affiliation[label4]{organization={Sydney Institute for Astronomy},
            addressline={School of Physics, The University of Sydney}, 
            city={Sydney},
            postcode={2006}, 
            state={NSW},
            country={Australia}}

\affiliation[label5]{organization={ARC Centre of Excellence for Gravitational Wave Discovery
(OzGrav)},
            city={Hawthorn},
            postcode={3122}, 
            state={Victoria},
            country={Australia}}

\begin{abstract}

Long Period Transients (LPTs) are a recently identified class of sources characterized by periodic radio bursts lasting seconds to minutes, with flux densities that might reach several tens of Jy. These radio bursts repeat with periodicity from minutes to hours, and they exhibit strong polarization and transient activity periods. To date, about 12 such sources have been identified, which might encompass the same or different physical scenarios. Proposed explanations include binary systems with a white dwarf and a low-mass star companion, slow-spinning magnetars, highly magnetized isolated white dwarfs, and other exotic objects. In a few cases the optical counterpart indeed points toward a white dwarf with a low-mass companion, while in other cases, transient X-ray emission was detected, very common in magnetars. However, despite being able to reproduce partially some of the characteristics of LPTs, all the proposed scenarios find difficulty in explaining the exact physical origin of their bright, highly polarized and periodic radio emission. We review here the state-of-the-art in the observations and interpretation of this puzzling class of radio transients.

\end{abstract}



\begin{keyword}

\end{keyword}

\end{frontmatter}



\section{Introduction}
\label{sec: intro}

\begin{figure*}[t]
\centering
\includegraphics[width=15cm,angle=0]{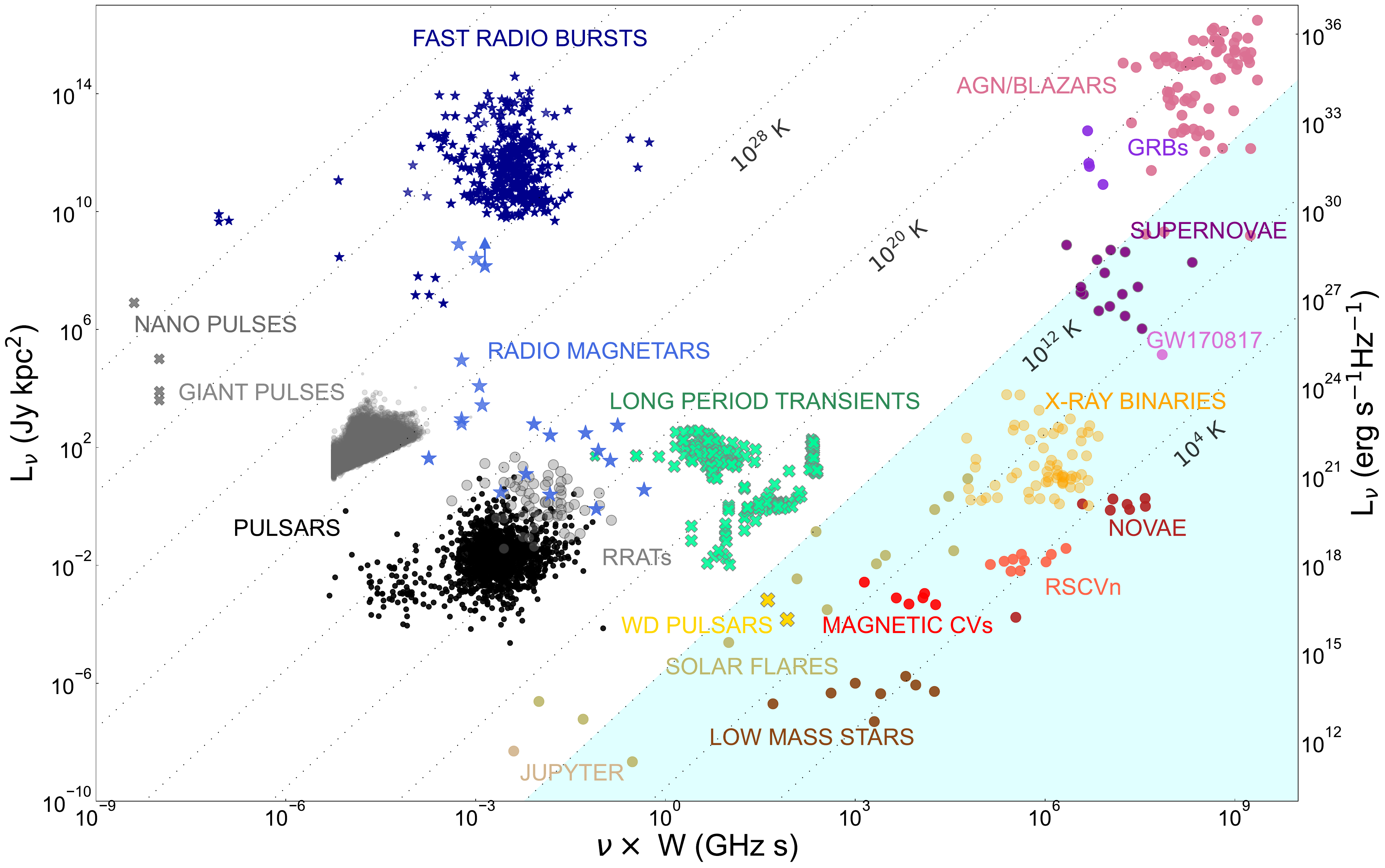}
\caption{Radio-transient plane including all transient sources. LPTs radio peaks are reported in green, and WD pulsars in yellow. The light blue region represent a parameter space for non-coherent radio emission. Data collected from: \citet{Keane2018,Nimmo2022, HurleyWalker2022, Rea2022, 2025Natur.642..583W}.}\label{fig:transient_plane}.
\end{figure*}

The transient radio sky has long been a fertile ground for serendipitous discovery. It includes a wide variety of sources, from nearby flare stars, pulsars, Fast Radio Bursts (FRBs), compact objects in binary systems, to distant gamma-ray bursts (GRBs) \citep{2025arXiv251110785M}. Fast, coherent emitters like pulsars and FRBs cluster at high radio luminosities and short durations. Slower, incoherent sources such as flare stars and active binaries populate the lower-luminosity, but longer-timescale regime. X-ray binaries exhibit radio flares from relativistic jets, often linked to accretion state changes. GRBs produce bright, decaying afterglows from synchrotron emission in external shocks, lasting days to weeks. Magnetars emit both short radio FRB-like bursts, bright single pulses and pulsar-like emission, despite some emission differences. A few white dwarf (WD) binary systems like AR Scorpii \citep{marsh_radio-pulsing_2016, 2017NatAs...1E..29B} and eRASS\,J1912-4410 \citep{Pelisoli2023} (also named WD pulsars) produce modulated radio pulses whose origin remain unclear, possibly due to the interaction of the WD magnetic field with the low-mass star companion wind \citep{2016ApJ...831L..10G}.
The radio transient plane provides a unifying framework for mapping the diversity and underlying physics of transient radio sources (see Figure\,\ref{fig:transient_plane}). In recent years, a new type of radio variable sources has emerged, referred to as Long Period Transients (LPTs; sometimes confusingly called in the literature: Long Period Pulsars --LPPs, Ultra-long period Pulsars -- ULPs, or Ultra-Long Period Magnetars -- ULPMs, etc.), characterized by bright, highly polarized radio pulses showing periodicities from minutes to hours (see Table\,\ref{tab:lpt_general}). Similarly highly polarized radio pulses are observed for pulsars and magnetars, but the long periodicities are in stark contrast to the fast millisecond to second-scale modulation typical of such sources. On the other hand, the radio emission from the two known WD pulsars is stable, generally with lower (and circular) polarization and a significantly lower radio luminosity.


The prototype of this newly discovered class is GLEAM$-$X\,J1627$-$5235, discovered by using the Murchison Widefield Array \citep[MWA;][]{2013PASA...30....7T}. 
This source emitted radio pulses with 95\,\% linear polarization and a periodicity of $\sim$18.2 minutes, persisting only over several months. Its coherent emission, high polarization and low-frequency spectrum, suggested a potential link in origin with neutron stars (NSs), yet
the long period challenged conventional models of rotation-powered pulsars and magnetars. After its discovery, a possible similarity to the a 77\,min periodic radio transient in the Galactic Center, GCRT\,J1745$-$3009 \citep{Hyman2005}, has been suggested.  
Subsequent discoveries, such as the 22\,minute periodic radio transient GPM\,J1839$-$10 \citep{2023Natur.619..487H}, have reinforced the idea that such long-period radio transients may represent either a previously unrecognized stage in strongly magnetized NS evolution or a fundamentally new class of radio emitting systems. However, these sources often defy categorization within the traditional pulsar-magnetar taxonomy, with some potentially violating the so-called ``death valley'' in the $P$-$\dot{P}$ diagram \citep{2023Natur.619..487H, 2024ApJ...961..214R}.

The growing sample of LPTs now comprises 12 sources ranging from $\sim$7\,min to $\sim9$\,hr periodicity, a few with a pulsar-like polarization, microstructures, and position-angle (PA) swings, others with an optical counterpart pointing to a WD in a low mass binary system, and in some cases having X-ray emission. Furthermore, some slow radio pulsars are also starting to be discovered with periods reaching even 76\,s \citep{2022NatAs...6..828C}, bridging the gap between the known pulsar-magnetar population and these LPTs, confusing further the final interpretation on the LPT nature. A conclusive definition of this new class remains premature, as it lies within a recently opened discovery space that is still being actively explored. For the time being, however, the working definition we adopt is of radio periodic emitters characterized by bright highly polarized radio pulses with periods ranging from minutes to several hours.

\begin{figure*}[t]
\centering
\includegraphics[width=15cm,angle=0]{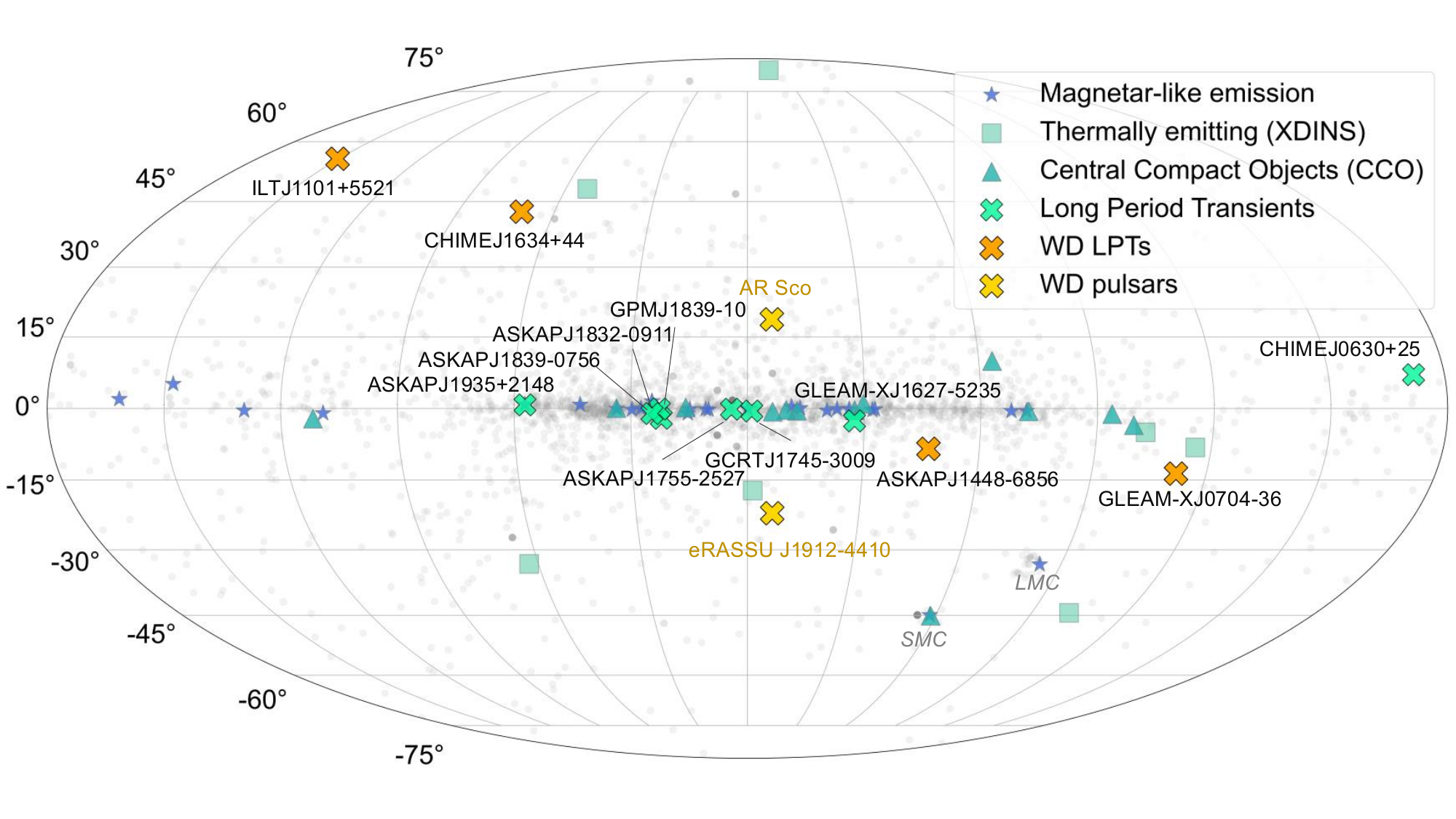}
\caption{Position distribution of LPTs compared to different source classes in Galactic coordinates. LPTs are represented by green crosses, LPTs associated with a WD system as orange crosses, and the two WD pulsars as yellow crosses (see also Table\,1.\label{fig:positions}). Grey dots are rotational powered pulsars from the ATNF Catalog\cite{Manchester2005}.}
\end{figure*}

This possibly heterogeneous class (see Figure\,\ref{fig:positions} and Table\,\ref{tab:lpt_general}) invites a re-evaluation of NS magnetospheres, magnetic field decay, binary white dwarf evolution and radio emission mechanisms. Moreover, these sources highlight the importance of wide-field, low-frequency radio surveys in uncovering rare, low-luminosity populations that evade detection in higher-frequency regimes.

In this review, we provide an overview of the current state of knowledge on LPTs as comprehensive as possible given the rapidly evolving field. We summarize observational characteristics, review theoretical interpretations, and highlight open questions and future prospects for this rapidly evolving field.

\begin{table*}[t]
\centering
\footnotesize{
\begin{tabular}{l|rrrlccr}
\hline
\hline
Source & $l$ & $b$ & $P_1$ & $\dot{P_1}$  &  $P_2$ & References \\ 
 & (deg) & (deg) & seconds (minutes) & ($10^{-11}$s/s) & (hr)  &  \\ 
\hline
GCRT\,J1745$-$3009  & 358.8911 & -0.5409 &  4620.72 (77.01) & -- &  -- & \cite{Hyman2005}\\ 
GLEAM-X\,J1627$-$5235 & 332.4646  & -2.6009  & 1091.17 (18.18)  & $<$ 120   &  -- & \cite{HurleyWalker2022} \\ 
GPM\,J1839$-$10 &  22.1526 & -2.0629 &  1318.19 (21.97) &  $<$0.036 & 8.75 & \cite{HurleyWalker2023} \\ 
ASKAP\,J1935$+$2148 & 57.1901  &  +0.7453 & 2225.31 (53.76) & $<$12 & -- & \cite{clk+24} \\ 
CHIME\,J0630$+$25 &  187.9709 &  +7.0606 & 421.35 (7.02) & 0.08  & -- & \cite{Dong2025} \\ 
ASKAP/DART\,J1832$-$0911 & 22.6406  & -0.0839 & 2656.24 (44.27)  & $<$0.9 & --  & \cite{2025Natur.642..583W}\\ 
ILT\,J1101$+$5521 & 150.4551  & +55.5199  & 7531.78 (125.53)   & $<$1.71 &  2.09 ($= P_1$) & \cite{2025NatAs...9..672D}\\ 
GLEAM$-$X\,J0704$-$36 & 247.8955  & -13.6352 & 10496.55 (174.94) & $<$3.9 &  2.91 ($= P_1$) & \cite{2024ApJ...976L..21H} \\ 
ASKAP\,J1839$-$0756 & 24.5450 & -1.0557 & 23221.70 (387.02)  & $<$16000 &  -- & \cite{2025NatAs...9..393L}\\ 
ASKAP\,J1448$-$6856 & 313.1644  & -8.4338 & 5631.07 (93.85) & $<$2200 & -- &  \cite{Anumarlapudi2025}\\ 
CHIME/ILT\,J1634$+$44 & 70.1692 &  +42.5754 & 841.24 (14.02) & --0.9 & 1.16  & \cite{2025ApJ...988L..29D}\\
ASKAP\,J1755$-$2527 &  4.116664 &  -0.122707 & 4186.32 (69.77) & $<$1 & --  & \cite{Dobie2024}\\
\hline
AR\,Sco & 0.11599  & 353.52  & 117.12 (1.95)  & 0.068 & 3.56  & \cite{marsh_radio-pulsing_2016}\\ 
eRASSU J1912$-$4410 & 353.3432  &  -22.0613 & 319.3 (5.32) & $<$0.6 & 4.03  &  \cite{Pelisoli2023}\\ 
\hline
\hline 
\end{tabular}}
\caption{General properties of LPTs and the two WD pulsars. Reported X-ray fluxes (0.5--10\,keV) and Optical/IR magnitudes are the observed ones, not corrected for absorption or extinction. $P_1$ refers to the first detected radio periodicity, while $P_2$ is the second discovered periodicity either in radio or optical. We refer to the section on each source for all due references. } \label{tab:lpt_general} 
\end{table*}

\section{Source sample}
\label{sec: sources}

In the past few years the number of LPTs have been growing enormously, encompassing some apparent similarities but also certain differences. In this section we summarize their observational characteristics.

\subsection{GCRT\,J1745$-$3009}
GCRT J1745$-$3009 is a periodic radio transient discovered by \citet{Hyman2005}, \textit{a posteriori} recognized as a possible LPT. First detected in archival 330-MHz Very Large Array (VLA) observations near the Galactic Center, it exhibited a series of $\sim$1-Jy bursts, each lasting approximately 10\,minutes and recurring every 77\,minutes over a span of several hours. The  brightness temperature ($\sim$10$^{16}$\,K) and energy density exceeded those observed in most other classes of radio transient sources, and they were consistent with coherent emission processes. No counterparts have been identified at X-ray, optical, or infrared wavelengths, complicating its classification. Additional activity was recorded in 2002 and 2003 \citep{Hyman2009}, though the source remains intermittent and unpredictable in its active phases.

Despite extensive follow-up, the lack of multi-wavelength detections \citep{Kaplan2008} and irregular activity cycles have kept GRT\,J1745$-$3009 an unsolved mystery for about two decades, when LPTs started to pop up.

\begin{figure*}[t]
\centering
\includegraphics[width=14cm,angle=0]{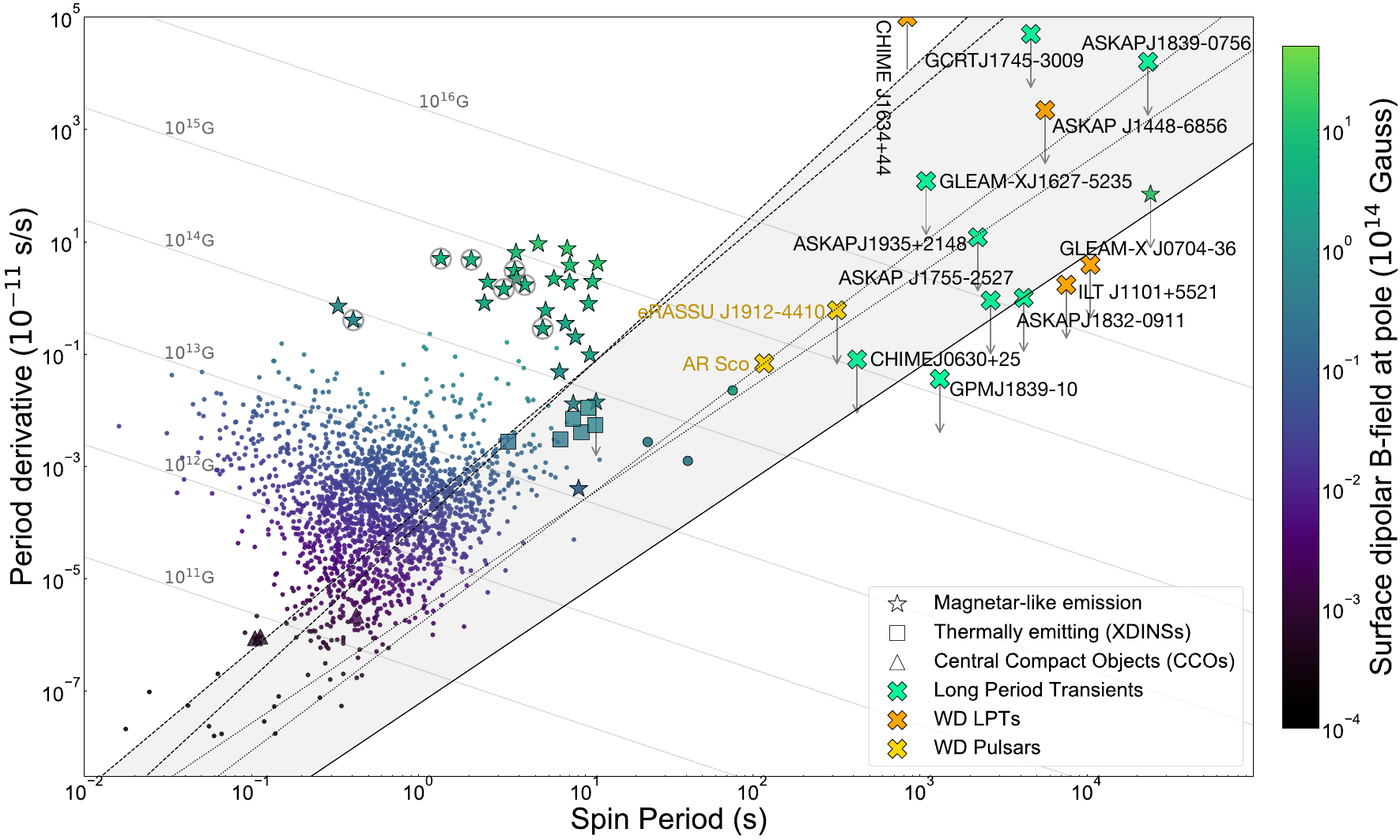}
\caption{Period ($P$) versus period derivative ($\dot{P}$) diagram for different pulsar classes, with LPTs' periodicity overplotted for comparison (note for most LPTs their $\dot{P}$ are probably not due solely to dipolar losses, as for pulsars).  We show isolated ATNF radio pulsars \citep{2005AJ....129.1993M} (dots), pulsars with magnetar-like X-ray emission \citep{ReaDeGrandis2025} (stars), including the long-period magnetar 1E~161348-5055 \citep{2006Sci...313..814D,2016ApJ...828L..13R}, X-ray Dim Isolated NSs (XDINSs; squares) and Central Compact Objects (CCOs; triangles). The three long-period radio pulsars are reported as larger circles \citep{Tan-etal2018,2022NatAs...6..828C,44sP_2025}. LPTs are represented with green crosses, the LPTs associated with a WD system as orange crosses, and the two WD pulsars as yellow crosses (their period derivatives are estimated from \cite{Pelisoli2022, Pelisoli2023}). Vertical arrows represent $\dot{P}$ upper limits, while only a vertical line is plotted for CHIME\,J1634$+$44 that has a negative $\dot{P}$ measurement.
The oblique lines represent different configurations of the pulsar death lines: the dashed lines correspond to the theoretical death-lines for a pure dipole, dotted lines for a twisted dipole, and the solid lines for the twisted multipole configuration \citep{1993ApJ...402..264C, 2000ApJ...531L.135Z, 2023Natur.619..487H, 2024ApJ...961..214R}. The shaded region is the death valley for pulsars.}\label{fig:p_pdot}
\end{figure*}

\subsection{GLEAM-X\,J1627$-$5235}
\begin{figure}[t]
\centering
\includegraphics[width=0.9\linewidth, height=7.8cm, angle=0]{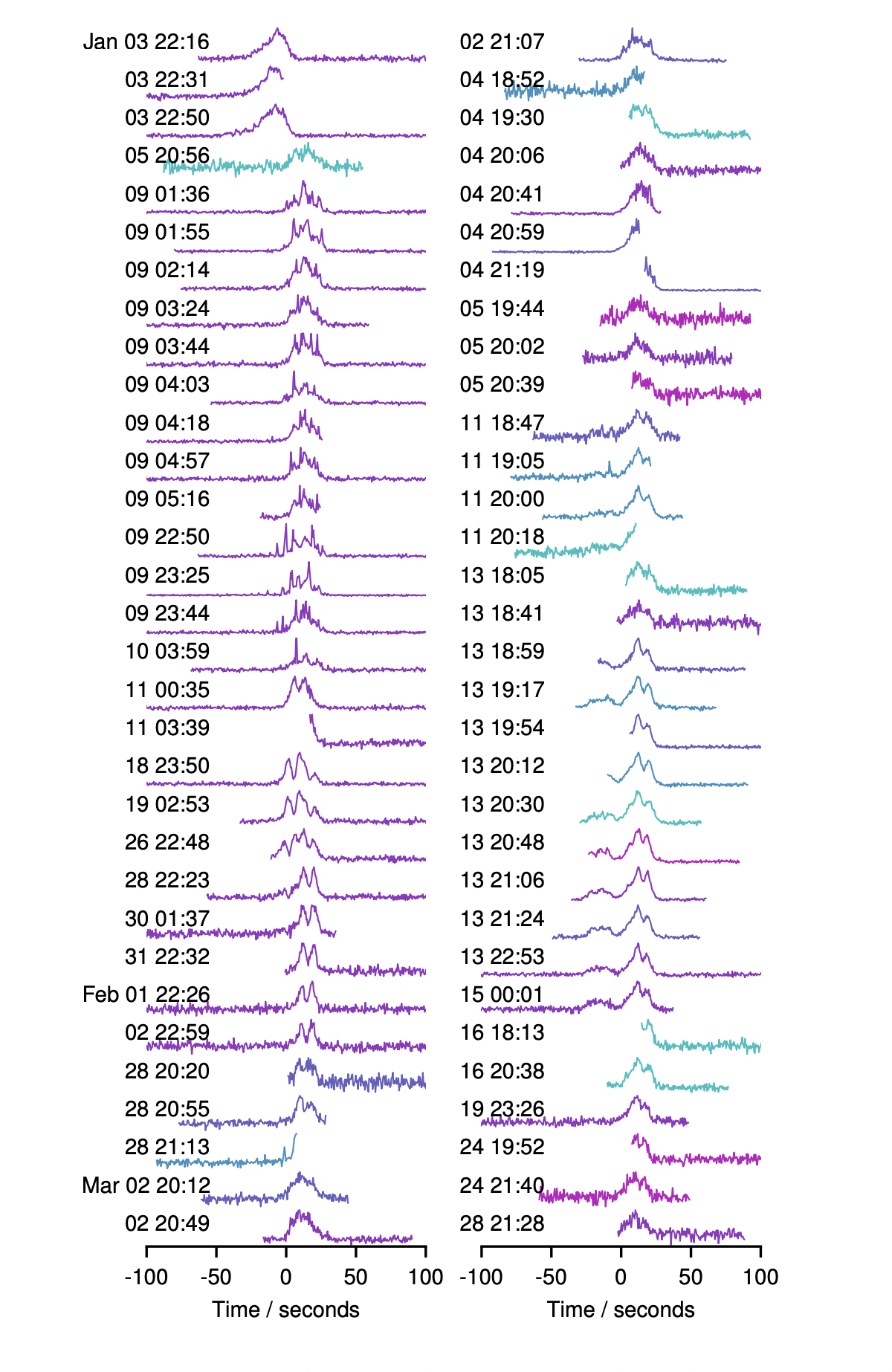}
\caption{Radio Pulses stack of data on the LPT GLEAM-X\,J1627$-$5235 (from \citealt{HurleyWalker2022}) .\label{fig:profiles}}
\end{figure}

GLEAM-X J1627$-$5235 was detected as a powerful radio transient discovered using archival data from MWA \citep{HurleyWalker2022}. Located near the Galactic plane, this source emitted highly polarized, coherent radio bursts with a periodicity of approximately 18\,min.
The periodic radio bursts, each lasting between 30 and 60 seconds, were detected across the 72–231 MHz band with strong linear polarization and a steep radio spectrum, implying a brightness temperature exceeding $10^{14}$\,K and thus again requiring a coherent emission process. The emission remained detectable for 3 months (January -- March 2018), with slow pulse evolution over this window, from a single leading pulse, to a double-pulse structure, to a single trailing pulse (see Figure\,\ref{fig:profiles}).  This transient radio activity initially pointed to a potential connection with magnetars that similarly show radio emission only during months-long outbursts.
No clear counterpart has been identified at X-ray, optical, or infrared wavelengths, despite extensive follow-up \citep{Rea2022,Lyman2025}. The dispersion measure and sky location suggest the source lies at a distance of roughly 1\,kpc, placing it within our Galaxy. The duration, regularity, and high luminosity of the bursts argued against typical stellar flaring activity from the very beginning, and instead pointing toward a compact, rotating object with a stable beam crossing Earth's line of sight.

\subsection{GPM\,J1839$-$10}
GPM J1839$-$10 was discovered using data from the MWA and confirmed with follow-up observations from MeerKAT, ASKAP, and the Parkes/Murriyang radio telescope. At odds with previously-discovered sources it shows continuous activity over more than three decades \citep{2023Natur.619..487H}, enabling long-term radio timing.
The source emits bright, linearly and circularly polarized, broadband radio pulses with a stable periodicity of $P = 1317.2 \pm 0.2$ seconds (approximately 22\,minutes). In combination with the limit on its period derivative ($<$\gpmPdot\,\Pdotunit), this source exceeds the traditional ``death line'' for pulsars (see Figure\ref{fig:p_pdot} and \S\ref{sec:interpretation}), where coherent radio emission was thought to be suppressed due to insufficient voltage drop across the magnetic polar cap \citep{2023Natur.619..487H, 2024ApJ...961..214R, Tong2023}.\\
The source is located within the Galactic plane (Figure\,\ref{fig:positions}), with a dispersion measure (DM) of $273.5 \pm 0.2$ pc cm$^{-3}$ implying a distance of roughly 6.1\,kpc based on the YMW16 Galactic electron density model. Significant polarization changes (see Figure\,\ref{fig:polarization}).and sub-pulses have been observed \citep{Men025}. Archival imaging data from 1988--2022 (including MOST, GMRT, VLA, and MWA) show no evidence for persistent radio emission, and deep observations with MeerKAT placed a 3$\sigma$ upper limit of $\sim$60\,$\mu$Jy at 850\,MHz. X-ray observations with XMM-Newton and Swift show no detectable X-ray counterpart, with 0.3–10 keV luminosity limits on the order of $L_X \lesssim 10^{32}$~erg~s$^{-1}$ assuming a thermal spectrum at 6\,kpc. No infrared or optical counterpart has been identified in VVV or Gaia DR3 data, with a potential (to be confimed) candidate from a deep IR GranTecan observation \citep{HurleyWalker2023}. Recently HiperCAM has detected a possible optical periodicity from a potential counterpart but further observations are needed to increase the detection significance \citep{Pelisoli2025}. \\
Unlike previous discoveries, and despite its decades-long activity, the source was seen to only intermittently produce pulses, being active for $\sim$30\% of the time. This was a mystery, until contiguous observing with ASKAP, MeerKAT, and the VLA for over 36 hours enabled a deep study of its pulse times of arrival (see Figure\,\ref{fig:aroundtheworldobs}). From this, \cite{2025arXiv250906315H} found evidence of an underlying orbital period of $\sim$8.75\,hr, well-modelled by a binary model comprising a WD and a low-mass star (see Figure\,\ref{fig:wdmodel}). 

\subsection{ASKAP\,J1935$+$2148}
ASKAP\,J1935$+$2148 was discovered serendipitously in a 6h long observation of GRB~221009A at 887.5 MHz \citep{clk+24}, repeating every 53.8\,min. Follow-up observations with the MeerKAT telescope revealed further, weaker pulses ($\sim$26$\times$ fainter) with drastically different properties compared to the ASKAP detections. Additionally, the MeerKAT observations allowed a DM estimate of $145.8\pm3.5$ pc\,cm$^{-3}$. Observationally, ASKAP\,J1935$+$2148 appears to exhibit three emission states: the strong pulse mode consisting of 15 bright, $10-50$ seconds wide, and $>90\%$ linearly polarised pulses as seen with ASKAP; the weak pulse mode characterized by 4 faint, $\sim370$~ms wide, and $>70\%$ circularly polarised pulses as seen with MeerKAT; the completely nulling or quiescent mode as seen with both telescopes \citep{clk+24}. Remarkably, the same three distinct emission states are observed in the intermittent pulsar PSR~J1107$-$5907 \citep{yws+14}, each characterized by unique pulse profiles, polarization properties, and at times, varying pulse profile intensities. The intricate interplay of magnetic fields, plasma flows, and the magnetospheric environment could lead to the emergence of these different modes. This state changes might point to the interpretation of this object as a possible highly magnetic NS, however the exact nature of this source is still under investigation, and no statistically significant multi-wavelength counterpart has yet been identified. 

\subsection{CHIME\,J0630$+$25}
CHIME\,J0630$+$25 was discovered using the CHIME/FRB and CHIME/Pulsar instruments, exhibiting a periodicity of $\sim421$ s \citep{Dong2025}. The timing solution included an abrupt increase in the frequency. This might be reminiscent of a pulsar-like glitch, which had a magnitude consistent with the magnitudes expected in known NSs, both pulsars and magnetars. \\
Polarization and dispersion analyses reveal an abnormally high rotation measure, RM $ = -347.8\pm0.6$\,rad\,m$^{-2}$, in contrast with the relatively modest dispersion measure DM $= 22 \pm 1$ pc cm$^{-3}$. There are no obvious structures (e.g., HII, H$\alpha$) that could give rise to large RM contributions \citep{Dong2025}. Instead, it seems plausible that a significant fraction of the observed RM is contributed by a magneto-ionic structure near the source, with the structure perhaps relating to the progenitor system. A massive companion contributing to the local RM is another scenario that could occur \citep{Dong2025}.

\begin{table*}[t]
\centering
\footnotesize{
\begin{tabular}{l|ccccccc}
\hline
\hline
Source & Radio Flux & Pulse Duty Cycle & Activity & Polarization & Spectral index & Distance \\ 
 & (mJy) &   & (months) & (total) & $\alpha$  & (kpc)   \\ 
\hline
GCRT\,J1745$-$3009  & 500-1000  & 15\% &  $<1$ & 60--100\% &  -(13.5--6.5) & --  \\ 
GLEAM-X\,J1627$-$5235 & 5000-40000 &  6\% &  $<3$ & 88\% & -1.16  &   1.3$\pm$0.5  \\ 
GPM\,J1839$-$10 &  100-10000 & 12\% & $>$420 & 10--100\%  &  -3.17  & 5.7$\pm$2.9 \\ 
ASKAP\,J1935$+$2148 & 9-234  & 0.01--1.5\%  & $<12$ & 90\% &  +0.4/-1.2 &  4.3  \\ 
CHIME\,J0630$+$25 & 0.4-1.9 & 0.06--2.3\% &  $<1$ & -- & -(5.1--0.05) &  0.170$^{+0.31}_{-0.10}$ \\ 
ASKAP/DART\,J1832$-$0911 &  60--1800 & 5--10\%  & $>12$ & 92\% & -1.5  &  4.5$\pm$1.5 \\ 
ILT\,J1101$+$5521 &  41--256 & 2\%  & $<1$ & 51\% & -4.1  & 0.50$\pm$0.15 \\ 
GLEAM-X\,J0704$-$36 & 25--100 & 0.5--2\% & $>$120 & 10--50\% & -6.2 & 1.5$\pm$0.5  \\ 
ASKAP\,J1839$-$0756 & 2--1200 & 1.4--3.1\%  & $>$1 & 30--90\% &  -(2.7--2.9)  & 4.0$\pm$1.2   \\ 
ASKAP J1448$-$6856 & 3--12  & 15--70\% & $>$13 & 35--100\%   & -2.5 &  --  \\ 
CHIME/ILT J1634$+$44 &  400--9000 & 1.2\%  &  $>$54  &  100\% & -- & 1--4.3 \\
ASKAP J1755$-$2527 & 1--3000 &   0.5-3\% & $>$21 & 25\% & -2.4 & --     \\
\hline
AR\,Sco & 6--12 & 20--30\% & --  & 10--40\%   &  -(0.4--0.8) & 0.116$\pm$0.016 & \\ 
eRASSU\,J1912$-$4410 & 4--10 & 5--7\% & -- & 40--70\% & -(4.4 -- 0.8) &  0.237$\pm$0.005 & \\ 
\hline
\hline 
\end{tabular}
\caption{Radio properties of LPTs and the two WD pulsars. Note that these values can vary substantially from pulse to pulse, they might be dependent on the observed frequency and bandwidth of the instruments that detected each pulses, hence should be taken as general references. Check the relative papers on \S\ref{sec: sources} for precise numbers}. \label{tab:lpt_radio}}
\end{table*}

\subsection{ASKAP/DART\,J1832$-$0911}
Discovered contemporaneously by both ASKAP and the DAocheng Radio Telescope \citep[DART;][]{2024cosp...45.1699Y}, ASKAP/DART\,J1832$-$0911 (hereafter ASKAP\,J1832$-$0911; \citealt{2025Natur.642..583W, Li2024}) is the most luminous LPT discovered so far. In phase synchrony, it produces up to $\sim10^{32}$\,erg~s$^{-1}$ radio and $\sim10^{33}$\,erg~s$^{-1}$ X-ray pulses every 44\,minutes \citep{2025Natur.642..583W}. The radio and X-ray luminosity are transient and somehow correlated. The source is sufficiently radio-bright that the Galactic neutral \textsc{Hi} absorption line has been detected against its pulsations; relying on its dispersion measure, the distance estimate is $4.5^{+1.2}_{-0.5}$\,kpc. This places it at a similar distance to, and apparently within the shell of a supernova remnant, SNR\,G22.7$-$0.2, and the $\gamma$-ray source 4FGL\,J1832.0$-$0913 error ellipse, but these have been ruled out as associations and are merely chance geometric coincidences in this crowded field. Despite a VLBI-derived position with mas-uncertainties, no clear IR or optical counterpart has been identified. Apparently switching on in October or November 2023, the radio pulsations are visible up to at least 5\,GHz, while having a spectral turnover that renders the source invisible at low ($\lesssim300$\,MHz) frequencies. 

\subsection{ILT\,J1101$+$5521}
ILT\,J1101$+$5521 was discovered with the International LO-Frequency ARray \citep[LOFAR;][]{2013A&A...556A...2V}, producing $\sim$minute-long linearly-polarized pulses with a periodicity of 2\,hours, in this case the source radio activity period was around 25\% of the total observing time \citep[][]{2025NatAs...9..672D}. At a high Galactic latitude with low extinction, optical observations determined that the radio pulses are produced by an M dwarf$+$WD binary system with an orbital period that matches within errors the observed radio periodicity, with pulses produced when the two stars are in conjunction (90$^\circ$ to the phase in which the other optically-identified LPT GLEAM-X\,J0704$-$37 produces pulsations). Similar to GLEAM-X\,J0704$-$37, also for ILT\,J1101$+$5521 its polar-like nature has been proposed under the assumption that the radio emission is modulated by the spin period of the WD, hence the system spin would be synchronized to the orbital period observed in the optical. However, the asynchronous polar and intermidiate polar possibility has not been rule out yet for either of these two LPTs.

\subsection{GLEAM-X\,J0704$-$36}
One of the few LPTs with a confirmed optical counterpart, GLEAM-X\,J0704$-$36 \citep{2024ApJ...976L..21H} was discovered in searches of archival data taken with the MWA \citep{2022PASA...39...35H,2024PASA...41...54R,2025PASA...42..129H}. It produces short (30--60\,s) $\sim40$\% linear and $\sim10$\% circularly polarised pulses every $\sim$2.9\,hours. Spectroscopic follow-up by \cite{Rodriguez2025} demonstrated that the system is an M-dwarf/WD binary, in which radio pulsations are produced at the ascending node of the orbit (when the M-dwarf is at maximum redshift and the WD is at maximum blueshift). Weak H-$\alpha$ emission was also observed, partly from the M-dwarf, but also with a brightening that appears coincident with the orbital phase of the M-dwarf, but moving at a higher velocity, possibly linked to the radio emission. The optical orbital period is similar to the radio pulse periodicity, possibly pointing to a polar-like system if the radio emission will be conclusively related to the spin period. 


\subsection{ASKAP\,J1839$-$0756}
ASKAP\,J1839$-$0756 boasts one of the longest known period of this class at 6.45 hours and was discovered with the ASKAP telescope in a 15-minute Rapid ASKAP Continuum Survey low-band (RACS-low) observation at 943.5 MHz \citep{2025NatAs...9..393L}. Radio pulses were observed also at half the period with weak interpulses. This behaviour might be typical of the rotation period of a NS, indicative of emission from both magnetic poles of the object, but a geometric effect in a binary cannot be excluded.
The slope of the polarization PA of the weaker interpulse is reversed compared to the brighter main pulse, a characteristic observed in orthogonally rotating (NS) pulsars. On average, the main pulses of ASKAP\,J1839$-$0756 are 80\% linearly polarized and 40\% circularly polarized, while the interpulses are 90\% linearly polarized with negligible circular polarization \citep{2025NatAs...9..393L}. The rotation measure of $214\pm1.3$ rad m$^{-2}$ is consistent with that of the  Galactic foreground, and the pulses show very pulsar-like microstructures. Follow-up observations with MeerKAT revealed microscale substructure which enabled a peak brightness temperature measurement of $\sim10^{20}$ K.
No statistically significant counterpart has been identified yet for ASKAP\,J1839$-$0756. The radio luminosity of ASKAP\,J1839$-$0756 has been observed to monotonically decline, which is consistent with that observed in the six currently known radio-loud magnetars, but also with other transient LPTs.

\subsection{ASKAP\,J1448$-$6856}
ASKAP\,J144834$-$685644 is a newly discovered LPT identified by the Australian Square Kilometre Array Pathfinder (ASKAP). It emits highly polarized radio bursts (polarization fraction ranging between 35--100\%) every 1.5 hours.
The source produces elliptically polarized, narrow-band radio pulses with a steep radio spectrum that fades above 1.5\,GHz. The duty cycle of the bursts is around 70\% with large intensity variations typical of other LPTs. What sets this object apart is its detection across radio, optical, and X-ray wavelengths, making it the only LPTs observed with such a broad spectrum. Its spectral energy distribution peaks in the near-ultraviolet, suggesting a hot, magnetized origin, possibly another case of a WD plus an M-star binary system \citep{Anumarlapudi2025}.

\subsection{CHIME/ILT\,J1634$+$44}
Discovered contemporaneously by teams using CHIME \citep{2025ApJ...988L..29D} and LOFAR \citep{Bloot2025}  (hereafter CHIME\,J1634$+$44), this LPT lies relatively nearby and at high Galactic latitude. It has a primary (likely spin) periodicity of 841\,s, but its burst arrival patterns are indicative of a secondary (likely orbital) 4206\,s period. Its 10-s duration pulses can be either 100\% linearly or 100\% circularly polarised, with a dependence that (with small number statistics) seems related to the secondary periodicity. In contrast to all other LPTS, these radio pulses arrive with a significantly negative period derivative $\dot{P}=‑9.03(0.11)\times10^{-12}$\,\Pdotunit. \cite{2025ApJ...988L..29D} argue that the system might be inspiralling a NS binary, while \cite{Bloot2025} note a marginal optical detection is more consistent with WD binary, possibly a WD/K-dwarf or WD/WD binary.

\subsection{ASKAP J1755$-$2527}
This source was initially detected as a single, linearly-polarised two-minute pulsation with the ASKAP radio telescope, which showed a very pulsar-like polarisation angle swing of 90$^\circ$ across the pulse \citep{Dobie2024}. ASKAP\,J1755$-$2527 was subsequently re-detected with the MWA during monitoring of the Galactic Plane in 2024, although not seen in similar data 2022 \citep{2025MNRAS.542..203M}. The source was found to repeat every 1.16\,h, and follow-up with MeerKAT showed a flat PA over pulse phase, more similar to other LPTs. Noting a decline in the source brightness over several months of monitoring, and its previous non-detectability, \cite{2025MNRAS.542..203M} conjectured that should this source be a binary with a spin-orbital resonance being near,
but not exactly, a small-integer ratio, this would create a aliased beating pattern on a time-scale of months.

\begin{figure}[t]
    \centering
    \includegraphics[width=7cm,angle=0]{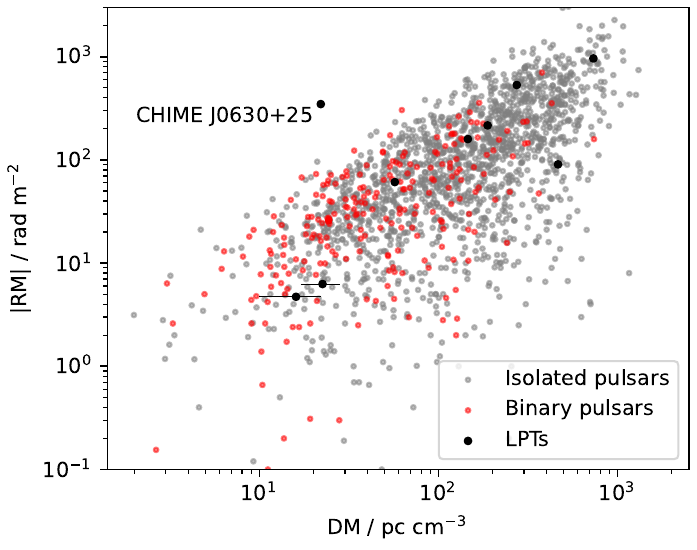}
    \caption{Absolute rotation measure (RM) against dispersion measure (DM) for the ATNF Pulsar Catalogue v2.6.5 \citep{2005AJ....129.1993M}, with grey and red points indicating, respectively, isolated and binary pulsars. The LPTs with known RMs and DMs are overplotted with black circles with 1-$\sigma$ error bars (for some sources, these are too small to be seen).}
    \label{fig:rmdm}
\end{figure}

\begin{figure}[t]
\centering
\includegraphics[width=\linewidth,angle=0]{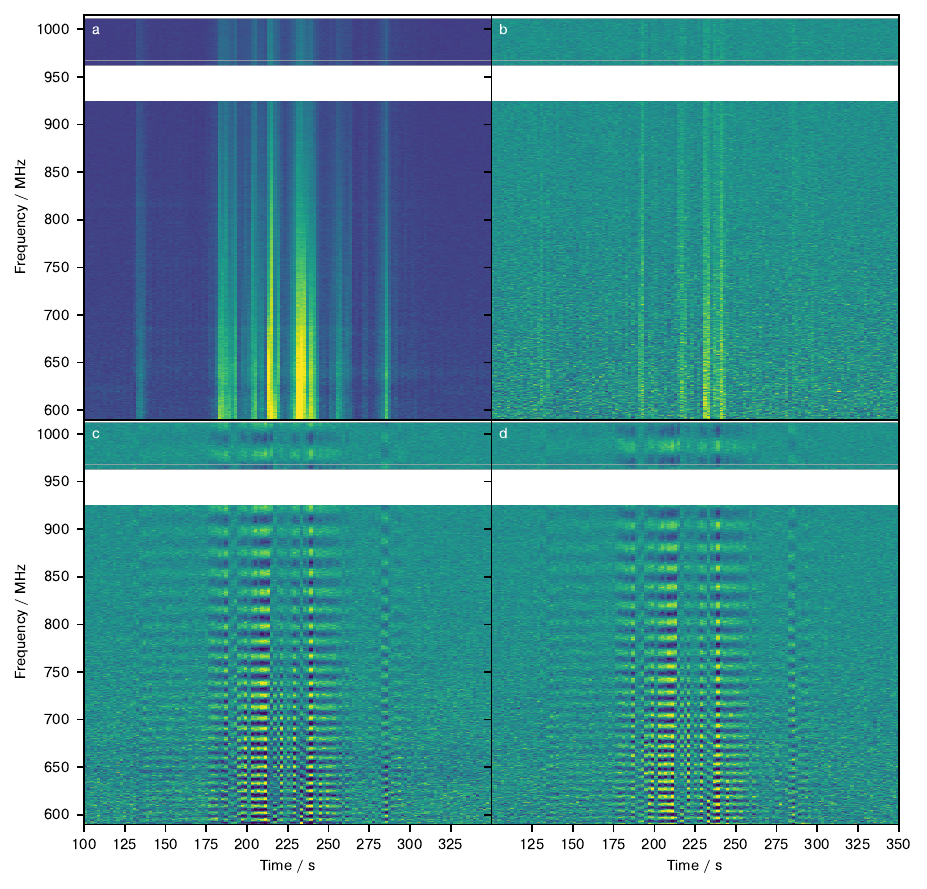}
\caption{Dynamic spectrum of a single pulse from GPM\,J1839$-10$. Reproduced from \cite{2023Natur.619..487H}, this plot shows the Stokes I, V, Q, and U intensity across respectively, panels a, b, c, and d. Significant ($\sim$10\%) circular polarisation is shown in Stokes V, and the linear polarisations demonstrate Faraday rotation consistent with a rotation measure of \gpmRM{}\,\RMunit{}. The polarisation angle is remarkably flat, except where orthogonal polarisation modes appear (e.g. at t=280\,s). }
\label{fig:polarization}
\end{figure}

\section{Radio properties}
\label{sec:radio_properties}

LPTs are characterized by the emission of coherent radio pulses lasting several seconds to minutes and repeating on unusually long periods of minutes to hours. These phenomena only became detectable with the advent of wide-field radio surveys capable of rapid imaging, sometimes referred to as ``fast imaging''. Prior to this, sources with such long periods were largely missed by traditional pulsar surveys, which are typically sensitive only to periods of a few seconds, as well as by conventional radio transient imaging programs. These LPTs are filling an under-populated region of the transient phase space, as seen in Figure\,\ref{fig:transient_plane}. Most LPTs have been detected at distances of several kiloparsecs and are distributed over a large range of Galactic latitudes, with a clear concentration near the Galactic plane (see Figure\,\ref{fig:positions}), although this distribution may in part reflect observational selection effects.
\begin{figure*}[t]
\centering
\includegraphics[width=14cm,angle=0]{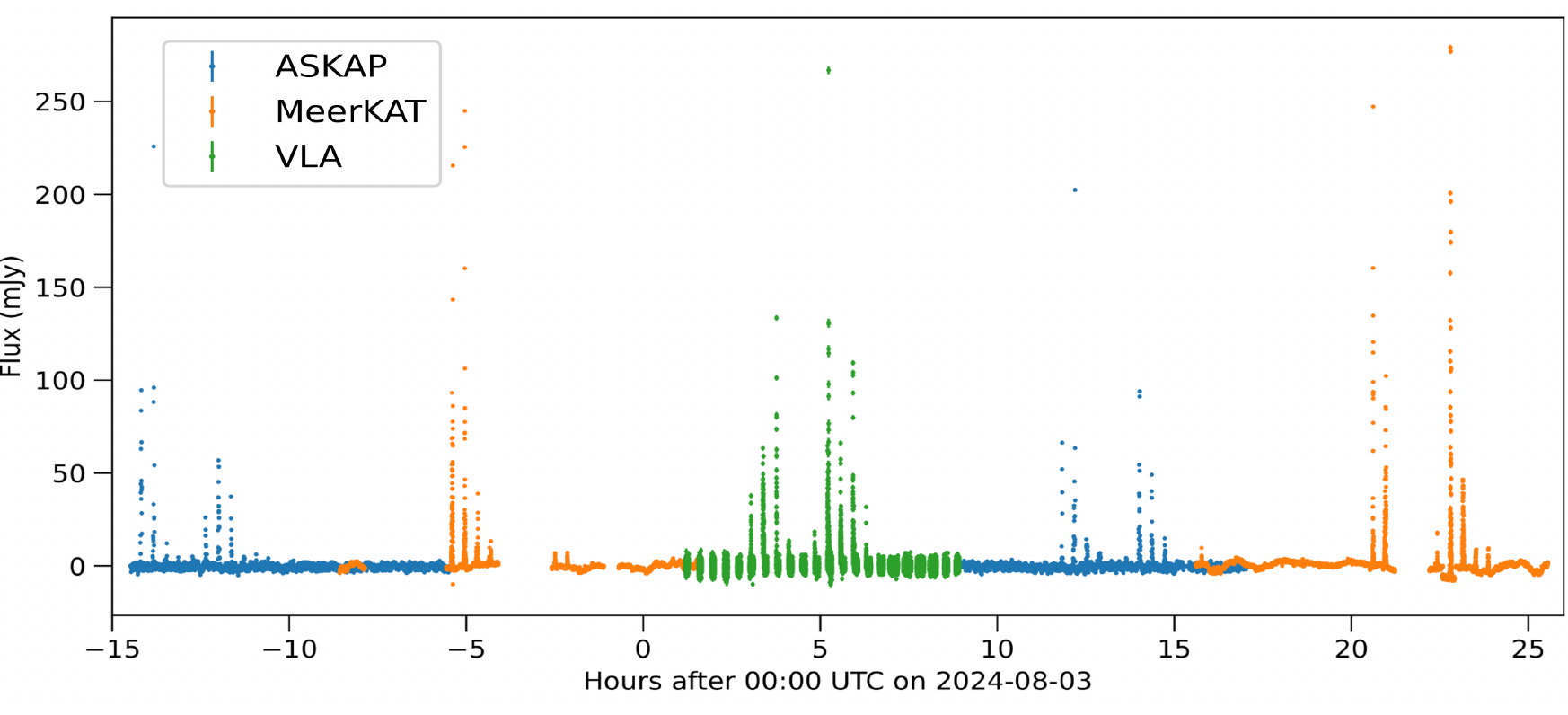}
\caption{Continuous radio observations lasting $\sim$40\,hours of GPM\,J1839$-$10 with ASKAP, MeerKAT and VLA (adapted from \cite{2025arXiv250906315H}).}
\label{fig:aroundtheworldobs}
\end{figure*}
In general, LPTs are characterized by highly polarized pulses, with the linear fraction often exceeding 70--90\% and the circular fraction reaching 20--30\%. These high polarization fractions are very common in coherent radio emitters like canonical pulsars and most repeating FRBs.
Some sources exhibit very flat polarization PA, while others (i.e. ASKAP\,J1839$-$0756; \citealt{2025NatAs...9..393L}) show swings and jumps similar to those observed in pulsars and some FRBs \citep{2025NatAs...9..393L, 2024ApJ...972L..20N}. In addition, GPM\,J1839$-$10 shows downward frequency drifting \citep{HurleyWalker2023, Men025}, a phenomenon usually associated with repeating FRBs and solar flares, which may indicate intrinsic effects in the magnetosphere or plasma lensing in the local environment. Some LPTs have been observed to be radio-loud only for a few weeks to months \citep[e.g.][]{HurleyWalker2022, clk+24}, whereas others have remained active for decades \citep{HurleyWalker2023}.

The episodic activity of LPTs may reflect magnetospheric instabilities, changes in particle acceleration regions, activation of coherent emission mechanisms that only operate intermittently, or potentially more than one progenitor type. Some LPTs exhibit distinct emission modes or states; a clear example is ASKAP\,J1935$+$2148 \citep{clk+24}, and in one case an interpulse has been detected (ASKAP\,J1839$-$0756; \citealt{2025NatAs...9..393L}). Even in sources without clear emission modes, individual pulses show large variability in both intensity and morphology across the class. These complex pulse morphologies often include substructure on timescales much shorter than the overall pulse duration, sometimes described as quasi-periodic modulations \citep{2025NatAs...9..393L, HurleyWalker2023}. They are theorized to arise from mechanisms related to magnetospheric radio emission or its propagation through the magnetosphere, and a range of timescales is often observed even within a given source. These features may be analogous to similar structures seen in magnetar bursts or high-magnetic-field pulsars, though their statistical significance is still under study \citep{2024NatAs...8..230K}.

LPTs pose significant challenges for conventional radio timing techniques due to their long periods, wide pulse widths, and intermittent emission, which make establishing phase connection between observations difficult. Additional complications arise from pulse-to-pulse variability, mode switching, and quasi-periodic substructure, all of which introduce timing noise and limit the precision of spin-period measurements. As a result, accurate determination of spin-down rates, potential orbital parameters, or long-term evolution of various properties requires high-cadence monitoring, often combining multiple telescopes and observing bands to fully characterize these systems.

In contrast to other highly intermittent radio source populations, i.e. the Rotating Radio Transients (RRATs; \citealt{mclaughlin_transient_2006}), spin-down energetics indicate that LPTs cannot be rotation-powered NSs and require other energy sources to explain the observed radio emission. For example, magnetic fields with a range of possible geometries may power the emission, whose luminosity is typically several orders of magnitude larger than the putative rotational budget (see also \S\ref{sec:radio_physics}).

A handful of LPTs have measured RMs, and their RM--DM relationship tends to be consistent with the general pulsar population (Figure~\ref{fig:rmdm}). The most notable exception, CHIME\,J0630$+$25, has RM $\approx -348$ rad m$^{-2}$, which is very high for its DM $= 22$ pc cm$^{-3}$ \citep{2025ApJ...988L..29D}. These values suggest that the source is embedded in locally magnetized plasma, either in a compact magnetosphere or in a surrounding nebula or binary environment. At present, RM variability over time and its correlation with pulse morphology, emission mode, or sub-pulse structure remain largely unexplored due to sparse timing and limited sample size. Nevertheless, RMs provide a useful diagnostic of the local magneto-ionic environment, and continued monitoring could reveal whether LPTs show RM evolution as seen in FRBs or remain relatively stable like ordinary pulsars.

\begin{table*}[t]
\centering
\footnotesize{
\begin{tabular}{l|cccr}
\hline
\hline
Source & X$-$ray flux & Opt/IR & Remarks & MW References \\ 
 & ($10^{-13}$erg/s/cm$^{2}$) & (mag)  &   \\ 
\hline
\hline
GCRT\,J1745$-$3009  &  $<$800  &  K$>$19.0  & Galactic center & \cite{Kaplan2008}\\ 
&  & & & \cite{Hyman2009}\\
\hline
GLEAM-X\,J1627$-$5235 & $<$0.03 & g$>$23.7  & Active for 3 months & \cite{Rea2022}\\ 
&  & & & \cite{Lyman2025} \\
\hline
GPM\,J1839$-$10 &  $<$0.08  &  $^*$K$_s$=19.7$\pm$0.2 & Active for 30 years & \cite{2023Natur.619..487H}\\ 
&  & & Binary system & \cite{Pelisoli2025}\\
\hline
ASKAP\,J1935$+$2148 & $<$0.02 &  $^*$K$_s$=17.1$\pm$0.1 & Mode switching & \cite{2022NatAs...6..828C} \\ 
&  & & Three emission states & \\
\hline
CHIME\,J0630$+$25 & $<$0.1 & -- & Close-by with High RM & \cite{Dong2025}\\ 
&  & & Possible presence of a glitch & \\
\hline
ASKAP/DART\,J1832$-$0911 &  2.3 ($<$0.1)  & K$>$19.8  & X-ray outburst and period & \cite{2025Natur.642..583W} \\ 
&  & &  & \cite{Li2024} \\
\hline
ILT\,J1101$+$5521 & $<$0.5 &   r=20.86$\pm$0.05 & WD$+$M$_{dw}$ (WD LPT) & \cite{2025NatAs...9..672D}\\ 
&  & & Optical radial velocity & \\
\hline
GLEAM-X\,J0704$-$36 & $<$0.2  &  G=20.79$\pm$0.01 & WD$+$M$_{dw}$ (WD LPT) & \cite{2024ApJ...976L..21H}\\ 
&  & & Optical radial velocity & \cite{Rodriguez2025} \\
\hline
ASKAP\,J1839$-$0756 & $<$2  &  K$>$19.6 & Interpulse & \cite{2025NatAs...9..393L}\\ 
&  & & & \\
\hline
ASKAP J1448$-$6856 & 0.26 & g=22.0$\pm$0.2  & WD$+$M$_{dw}$ (WD LPT) & \cite{Anumarlapudi2025}\\ 
&  & & Radio/optical/X-ray detection & \\
\hline
CHIME/ILT J1634$+$44 & $<$1.2 & g=25.3$\pm$0.4  & (WD LPT) & \cite{Bloot2025} \\
&  & & Possible Spin-up & \cite{2025ApJ...988L..29D}\\
\hline
ASKAP J1755$-$2527 & $<$0.6 &  g$>$24.5 &  & \cite{Dobie2024}\\
&  & & & \cite{2025MNRAS.542..203M}\\
\hline
\hline
AR\,Sco & 20  & g=14.4$\pm$0.2  & (WD Pulsar) & \cite{2017NatAs...1E..29B} \\ 
&  & & X-ray period & \cite{Takata2021}\\
\hline
eRASSU\,J1912$-$4410 & 1.66 &  G=17.09$\pm$0.02 &  (WD Pulsar) & \cite{Pelisoli2024}\\ 
&  & & X-ray period & \cite{Schwope2023}\\

\hline
\hline 
\end{tabular}
\caption{Multi-wavelength detections and upper limits for LPTs and WD pulsars (see Figure\,\ref{fig:xray}), plus additional remarks on each source. We refer to the section on each source for all due references. $^*$ Associated counterpart needs further confirmation.}\label{tab:mw}}
\end{table*}

\section{Synergies with X-ray and optical/IR emission}
\label{sec:x_opt_emission}
LPTs are by definition characterized by periodic radio bursts on minutes-to-hours timescales, often present during months-long radio-active periods and exhibiting highly variable radio fluxes (see Table\,\ref{tab:lpt_radio}). All LPTs have been searched for multi-band emission (see Table\,\ref{tab:mw}), typically in X-rays, optical and/or infrared bands. In most cases only upper limits could be derived, likely due to distance or source crowding in the region. However, X-ray emission has been observed in at least two cases.

\begin{figure}[t]
\centering
\includegraphics[height=6cm, width=7cm,angle=0]{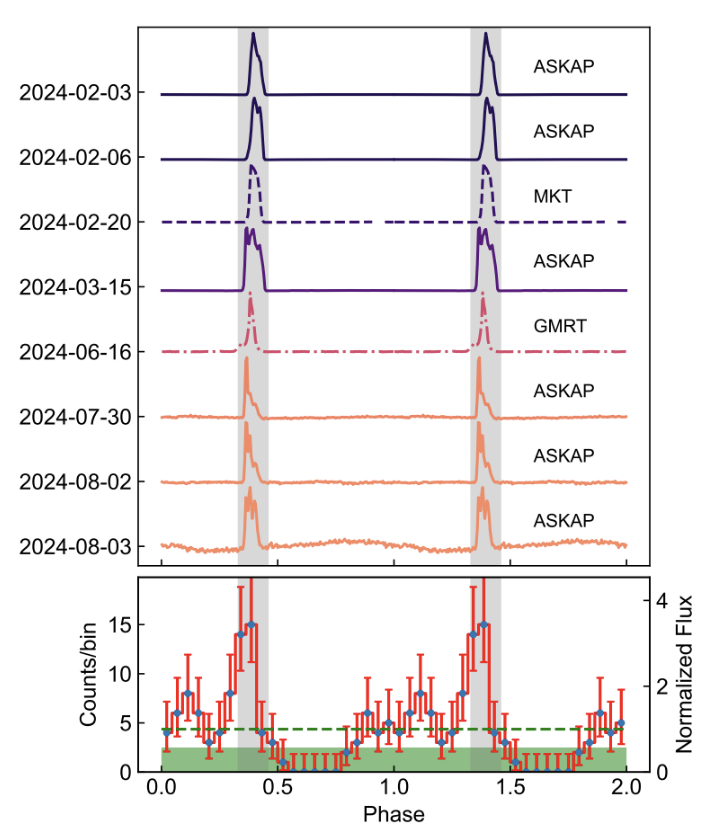}
\caption{Radio pulses (top 8 curves) versus X-ray folded lightcurve (bottom panel) for ASKAP\,J1832$-$0911. From \citealt{2025Natur.642..583W}.}
\label{fig:xrayprofile}
\end{figure}

The first discovered X-ray-emitting LPT is ASKAP\,J1832$-$0911 \citep{2025Natur.642..583W}, which not only showed an X-ray counterpart with a luminosity of $\sim10^{33}$\,erg~s$^{-1}$ (detected serendipitously by {\em Chandra}), but for which deep {\em XMM-Newton}, {\em Einstein Probe} and {\em Swift} archival observations showed that the source is extremely variable in X-rays. This was the first hint that the radio and X-ray activation periods might be correlated in LPTs, as they certainly are in ASKAP\,J1832$-$0911. Furthermore, timing analysis of the X-ray data showed a significant periodic variability at the same 44.2\,min radio period (see Figure\,\ref{fig:xrayprofile}), with an $\sim$100\% modulation fraction. The X-ray luminosity during the outburst exceeded the radio luminosity by about one order of magnitude, and, if the periodicity is interpreted as a spin period, it would exceed the rotational energy budget by more than five orders of magnitude. It is therefore impossible for the radio and X-ray emission to be powered solely by rotation, as in a typical isolated radio pulsar. The exact physics of the X-ray emission remains unclear. In magnetars, additional magnetic energy might power such transient X-ray emission; however, the deep upper limits in quiescence do not point to a strong surface magnetic field (unless invoking extreme configurations). On the other hand, X-ray outbursts are common in cataclysmic variables (CV) due to accretion episodes, but they are not seen in typical radio-emitting WD pulsars. However, both these systems do show steady and pulsed X-ray emission \citep{Takata2021, Pelisoli2023, Schwope2023}.  \\

The second X-ray-emitting LPT is ASKAP\,J1448$-$6856 \citep{Anumarlapudi2025}, showing a luminosity of $\sim10^{30}$\,erg~s$^{-1}$ (caveat the large distance uncertainties), but no variability or periodicity could be conclusively detected given the sparse and short X-ray observations. Also in this case, the X-ray emission is about three orders of magnitude brighter than the potential rotational power of this object.\\

All other LPTs have shallow X-ray observations, resulting in (not very constraining) upper limits ranging between $L_X < 10^{30}$--$10^{34}$\,erg~s$^{-1}$ depending on the source. Given the large X-ray variability observed in ASKAP\,J1832$-$0911, a single X-ray observation might have caught the source in a low-emission state and thus cannot constrain its possible X-ray outburst behaviour. One exception is GLEAM$-$X\,J1627$-$5235, which was observed very deeply with {\em Chandra}, reaching a luminosity limit of $L_X < 10^{29}$--$10^{30}$\,erg~s$^{-1}$. This deep upper limit in quiescence could exclude the presence of a typical magnetar in this source \citep{Rea2022}.

Optical/IR emission has been proposed for six LPTs. The associations of two of them, GLEAM-X\,J0704$-$36 \citep{2024ApJ...976L..21H, Rodriguez2025} and ILT\,J1101$+$5521 \citep{2025NatAs...9..672D}, have been confirmed via the detection of optical periodicity through radial-velocity measurements, consistent with the radio periods. In these two cases, their optical spectral energy distributions can be successfully modelled by a WD plus a low-mass star.

GLEAM$-$X\,J0704$-$36's optical spectrum is well modeled by a binary comprised of a relatively cold massive white dwarf (T$_{\rm eff} \sim7300$\,K, M$_{\rm WD} \sim 0.8-1.0$\,M$_{\odot}$) and an M dwarf (T$_{\rm eff} \sim3000$\,K, M$ \sim 0.14$\,M$_{\odot}$). Radio bursts in the system arrive when the WD is at nearly maximum blueshift and the M dwarf at nearly maximum redshift (namely the so-called ascending node). \\
ILT\,J1101$+$5521's optical spectrum is also well modeled by a binary comprised of a relatively cold white dwarf (T$_{\rm eff} \sim5500$\,K, M$_{\rm WD} \sim 0.77$\,M$_{\odot}$) and an M dwarf. Also in this case, the radio periodicity equals the optical orbit measured via radial velocity, but the bursts arrive when the M dwarf is at superior conjunction with respect to the line of sight. \\
Two other LPTs have confirmed optical/IR counterparts. ASKAP\,J1448$-$6856 showed a spectrum well modelled by a WD with a low-mass companion star and considerable optical variability, including an hr-scale optical flare \citep{Anumarlapudi2025}. This could either be due to a low-mass star flaring episode, or to a possible orbital variation. Its spectral energy distribution is less constrained than the other two, and no optical orbital variability could be detected due to limited data. \\
CHIME\,J1634$+$44 has a possible optical and UV detection using GALEX and UNIONS \citep{Bloot2025}, pointing to a possible WD in the system. The associations of the other two counterparts are less stringent: within the radio position of ASKAP\,J1935$+$2148 and GPM\,J1839$-$10 there are two IR sources which are possible candidates to be confirmed.

\begin{figure*}[t]
\centering
\includegraphics[width=15cm,angle=0]{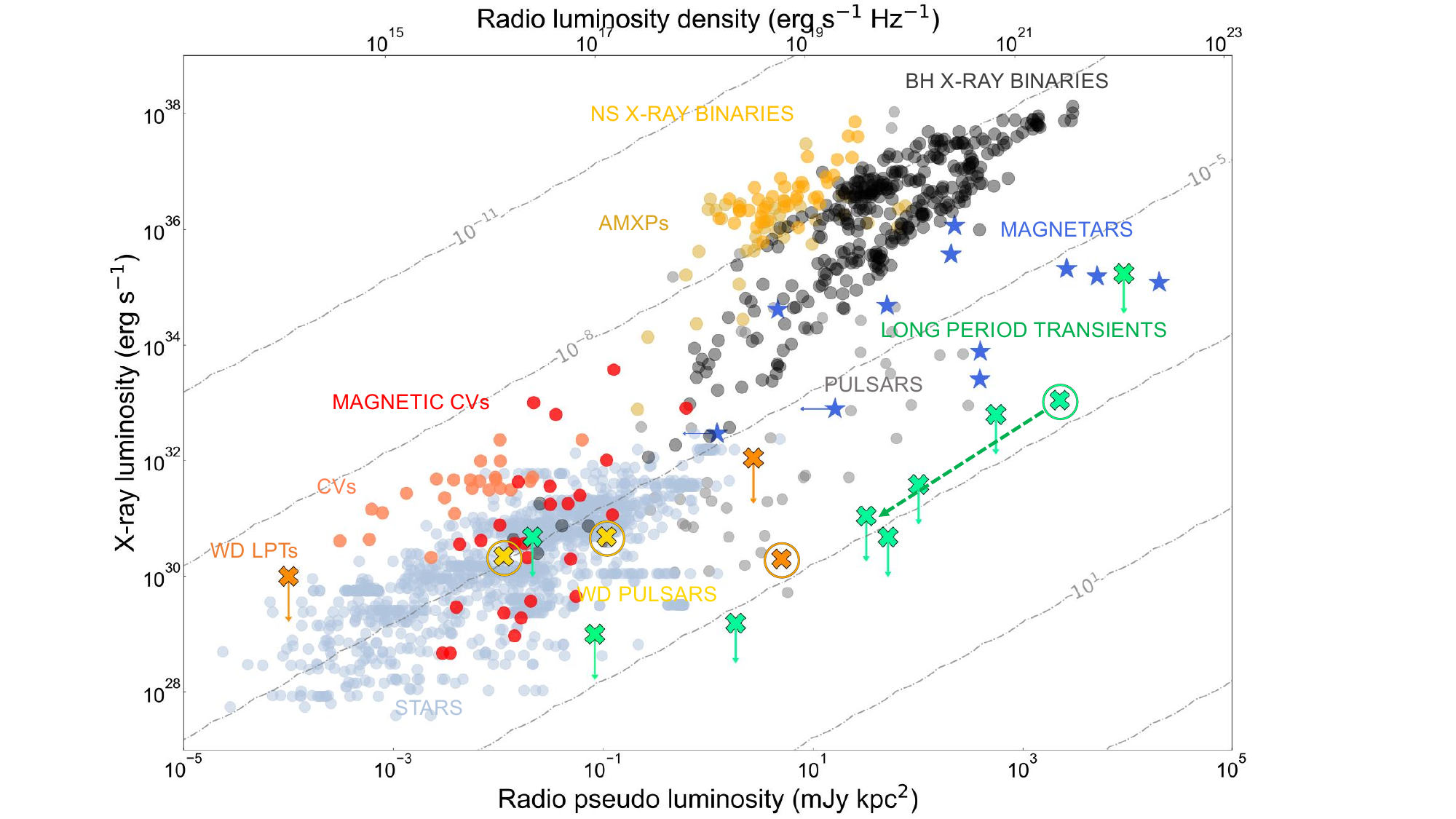}
\caption{Radio luminosity density versus X-ray luminosity for different classes of sources. Crosses represent LPTs and WD pulsars, and we label as WD LPTs those with a confirmed WD hosted by the system. Circles highlight X-ray detections: from left to right of the WD pulsars AR\,Sco and eRASSU\,J1912$-$4410, the WD LPTs ASKAP\,J1448$-$6856, and the green line indicates the X-ray outburst of ASKAP/DART\,J1832$-$0911. Adapted from \cite{Anumarlapudi2025} and \cite{2025Natur.642..583W}.}
\label{fig:xray}
\end{figure*}

\section{Source models}
\label{sec:interpretation}

\subsection{Binary white dwarfs with a low-mass star}
\label{sec:wd_binaries}
The detection of at least two confirmed WD binaries among LPTs, GLEAM-X\,J0704$-$36 and ILT\,J1101$+$5521, has by now proven their strong connection with this class of sources. This connection was proposed from the very beginning \citep{2023Natur.619..487H, 2025ApJ...981...34Q}, motivated by the similarity with the two radio-emitting WD binary pulsars AR~Sco and J1912$-$4410, despite the latter having lower radio luminosities and polarization fractions.

Binary systems hosting a WD and a low-mass main-sequence star are mainly observed in two different evolutionary phases. They originate from a common-envelope (CE) episode \citep{Paczynski1976,Ivanova2013}, in which the initially more massive star in a binary leaves the main sequence and expands into a giant, engulfing its companion. Dynamical friction within the CE extracts orbital energy and angular momentum, drastically reducing the separation between the two stars. The outcome of this interaction is a tight binary composed of a WD and a low-mass secondary star. Such post-CE binaries (PCEBs) are usually observed in a detached configuration \citep[e.g.][]{Rebassa2012}. Subsequent evolution is governed by angular momentum losses: magnetic braking driven by the wind of the secondary \citep{Verbunt1981}, and tidal forces that eventually synchronize the secondary’s rotation with the orbital period \citep{Fleming2019}. As magnetic braking gradually decreases the orbital separation, the donor eventually fills its Roche lobe, initiating mass transfer onto the white dwarf and marking the transition into the CV phase, where mass accretion episodes are expected \cite{Knigge2011}, and $\sim$35\% of these systems host magnetic WDs \cite{Pala2020}.

Systems like the WD pulsar AR~Sco are indeed in the detached phase, when the companion star is not filling its Roche lobe. A similar evolutionary stage may apply to GLEAM-X\,J0704$-$36 and ILT\,J1101$+$5521. At variance with the other WD radio binaries, assuming that the radio periodic emission is associated with the spin, these LPTs might have already reached synchronization (an assumption that still needs to be investigated).

A recent geometrical model by \cite{2025arXiv250906315H} succeeds in modelling the arrival times of the radio pulses of the most active LPT, GPM\,J1839$-$10, deriving an orbital period of $\sim$8.75\,hr and identifying the radio pulse periodicity as the beat frequency between the orbital and spin periods (see Figure\,\ref{fig:aroundtheworldobs}). This system is also composed of a WD and a low-mass star, and the model suggests that radio emission is triggered when the magnetic pole of the rotating WD intersects the companion’s wind in the binary orbital plane (see Figure~\ref{fig:wdmodel}). A similar emission geometry has been observed in WD pulsar systems such as AR~Sco and J1912$-$4410.

\subsubsection{Evolutionary link for LPTs, WD pulsars, and CVs}

\begin{figure*}[t]
\centering
\includegraphics[width=15cm,angle=0]{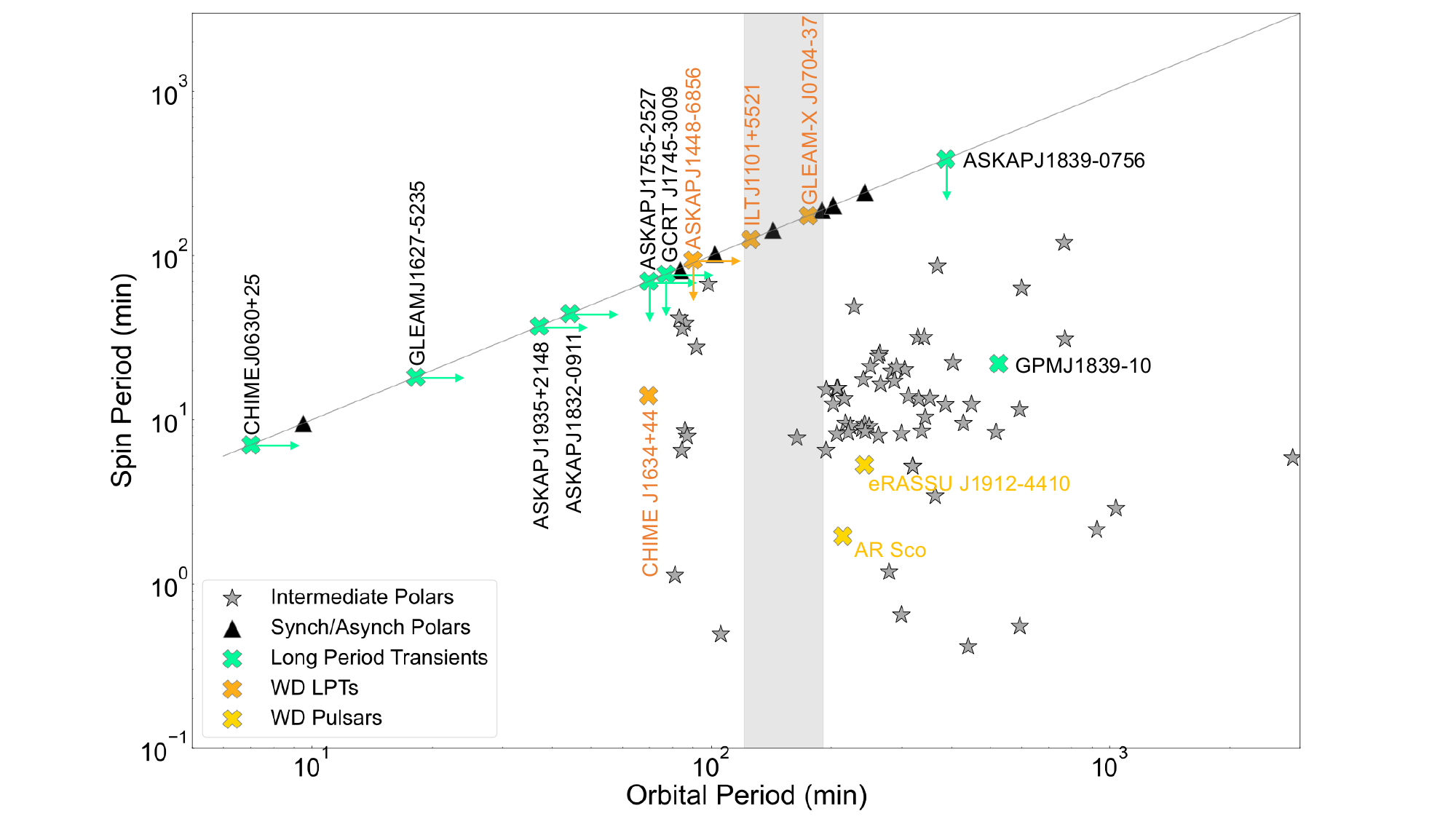}
\caption{Distribution of periodicities for LPTs, WD pulsars, a few known polars, and intermediate polars under the WD binary scenario (data taken from Koji Mukai online repository for IPs (see link in the GitHub LPT Review repository) and \cite{Ferrario2020}). The grey shaded region represents the CV ``period gap''. For most LPTs, the nature of the radio periods is unknown, and the arrows indicate their probable location if the systems are WD binaries.}
\label{fig:wdperiods}
\end{figure*}

The comparison between the periodicities of LPTs and CVs has opened an interesting discussion on the evolutionary connection between WD binary classes. CVs generally have orbital periods of $\sim$78\,min--10\,hr, with a debated ``period gap'' around 2.2--3.2\,hr (see the grey region in Figure~\ref{fig:wdperiods}), attributed to reduced magnetic braking when the donor becomes fully convective. This interpretation is consistent with the higher incidence of detached WD$+$M-dwarf binaries in this range. The location of both GLEAM-X\,J0704$-$36 and ILT\,J1101$+$5521 within the CV period gap might suggest that detached magnetic WD binaries can exhibit diverse radio phenomena (see also \cite{Coppejans2020} for a review on CV radio emission). However, the measured surface temperatures of the WDs hosted in these two LPTs are between $\sim$3000--5000\,K, which is relatively cold for a WD that has experienced accretion episodes (usually $>$10,000\,K). This points toward a pre-CV post–common-envelope system rather than a previously accreting CV, favouring a scenario of long-term cooling after envelope ejection (several Gyr) rather than within the limited $\sim$1\,Gyr gap-crossing timescale of CVs \citep{Rodriguez2025, CastroSegura2025, Yang2025arXiv250909224Y}.

Note that the shorter end of the LPT period distribution in this scenario should necessarily correspond to the spin period (or beat period) of the WD (see Figure~\ref{fig:wdperiods}). Standard binary evolution models predict that when the progenitor star loses mass and becomes degenerate, its radius begins to increase, causing the binary orbit to widen again (the so-called ``period bounce'' phenomenon). It is therefore plausible that, in these short-period LPTs, the longer orbital modulation is currently undetected, as was shown to be the case for GPM\,J1839$-$10, which has a well-measured $\sim$8.75\,hr orbital period \citep{2025arXiv250906315H}. Depending on the geometry of the systems and the typical duration of radio observations, it may not be possible to determine orbital modulation for all LPTs.

\subsection{Magnetar scenario}
\label{sec:magnetar_scenario}

Magnetars represent the most strongly magnetized end of the NS population, characterized by extremely strong magnetic fields, typically ranging between $10^{13}$ and $10^{15}$\,G. While they are primarily known for their high-energy X-ray and gamma-ray emission, a small number have also been observed to emit pulsed radio emission, often transiently and in connection with episodes of X-ray activity. This radio emission differs markedly from that of ordinary pulsars. Magnetar radio pulses tend to be highly variable in both flux and morphology, with pulse profiles that can change dramatically on timescales of minutes to hours, exhibiting complex substructures and strong polarization. Spectrally, magnetars deviate from typical pulsars, as their radio spectra are often flat or even inverted, allowing the emission to remain strong or increase at higher radio frequencies. However, magnetar radio spectra can also vary substantially, and the radio emission can cease abruptly and unpredictably, sometimes resuming after long intervals of quiescence \citep[for recent reviews]{2017ARA&A..55..261K, Esposito2021, ReaDeGrandis2025}. These properties made magnetars plausible candidates for LPTs from the outset \citep{HurleyWalker2022, Konar2023, Tong2023}. The challenge, however, lies in their long spin periods compared with the typical 0.3--12\,s range of magnetars, which requires additional investigation.

\begin{figure}[t]
\includegraphics[width=7.8cm,angle=0]{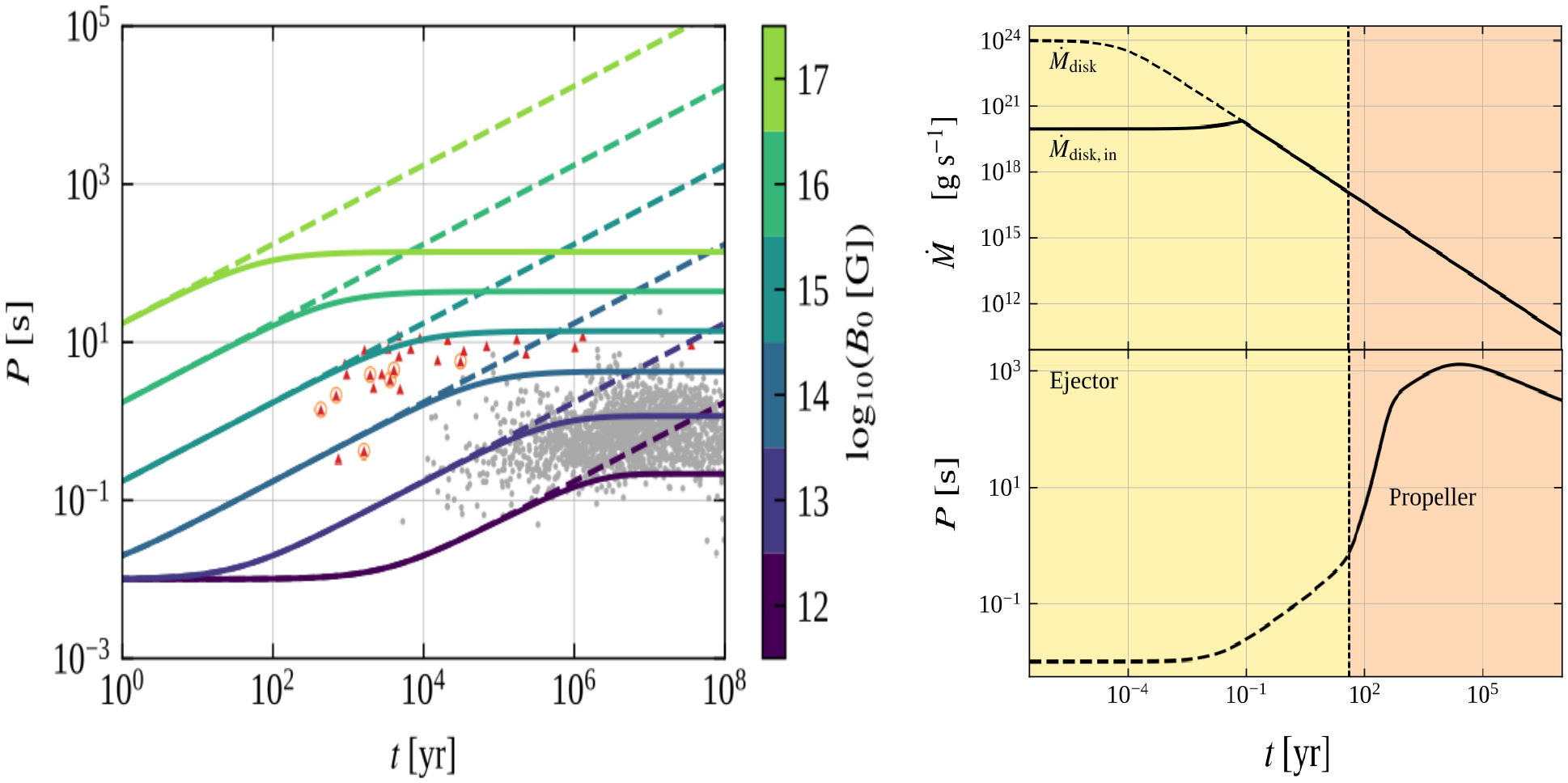}
\caption{Left panel: pulsar spin-period evolution over time for dipolar spin-down with different initial magnetic field strengths $B_0$. Dashed lines represent evolution with constant magnetic field, while solid lines show decaying-field models. Grey points indicate observed radio pulsars and red triangles indicate detected magnetars. Right panel: example of pulsar spin-down in the presence of a fallback disk. See \cite{2022ApJ...934..184R}.}
\label{fig:fallback}
\end{figure}

\begin{figure*}[t]
\centering
\includegraphics[width=13cm,angle=0]{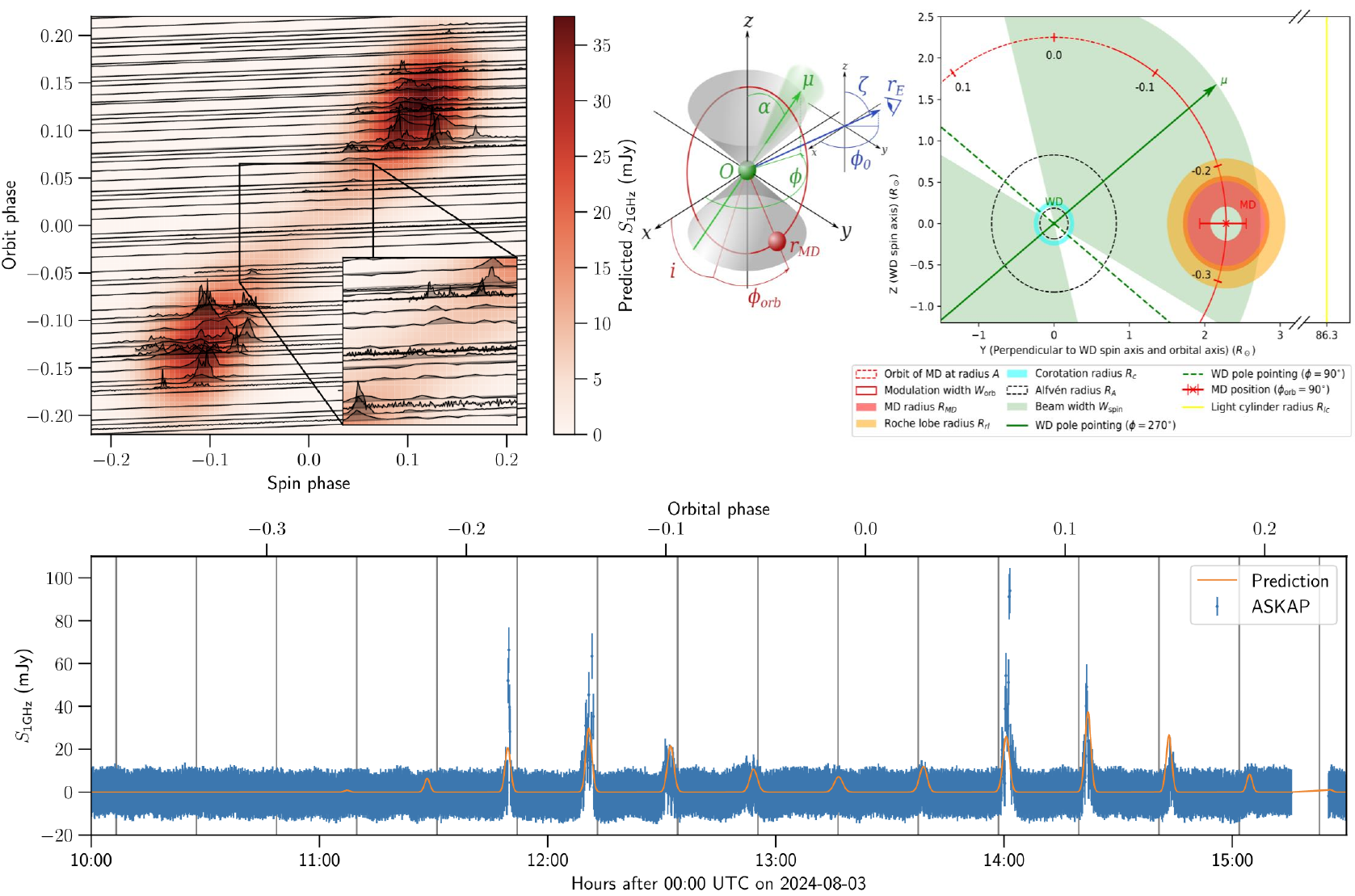}
\caption{Binary geometrical model fitted to the $\sim$40\,hours of continuous radio data of GPM\,J1839$-$10, constraining its binary parameters \citep{2025arXiv250906315H}.
\label{fig:wdmodel}}
\end{figure*}

The spin evolution of an isolated NS is governed primarily by magnetic dipole braking and field decay, leading to a gradual increase in period over time. Despite this spin-down, a pulsar or magnetar born with a typical period of $\sim$0.01--1\,s (see Figure~\ref{fig:fallback}, left panel) cannot slow down to more than about a hundred seconds through these processes alone. This limit is set mainly by the efficiency of magnetic field decay, which depends on the crustal resistivity \citep{Pons2013, 2022ApJ...934..184R}. However, fallback accretion after the supernova explosion may provide an additional spin-down channel. Under certain conditions, the newly formed NS could undergo substantial braking (see Figure~\ref{fig:fallback}). Simulations of magnetar spin evolution in the presence of a fallback disk have demonstrated the feasibility of producing young, slowly rotating magnetars \citep{2022ApJ...934..184R, Gencali2022, Fan2024, Xu2024}, although alternative scenarios based on accretion from the ISM have also been proposed \citep{Ferrario2025,Afonina2024}.

A magnetar-like outburst has been discovered in the $\sim$2\,kyr-old NS 1E~161348$-$5055, a Central Compact Object (CCO) located in the supernova remnant RCW~103 \citep{1E1613_Discovery, 2006Sci...313..814D,2016ApJ...828L..13R}. This thermally emitting X-ray source shows a periodicity of 6.67\,hr and no evidence for binarity, and is believed to be a magnetar slowed down by supernova fallback accretion. However, 1E~161348$-$5055 is apparently radio-quiet and emits predominantly in the X-ray band, with a possible infrared counterpart \citep{1E1613_IRcounterpart}.

On one hand, the magnetar scenario is attractive for explaining the radio properties of LPTs, as well as the X-ray/radio correlation observed in ASKAP\,J1832$-$0911. By invoking fallback accretion, it also provides a pathway to unusually long spin periods. On the other hand, this interpretation faces major challenges, particularly the low quiescent X-ray luminosities observed in LPTs (see e.g. \citealt{Rea2022}) and the unexpectedly large number of detected sources, which would imply an unrealistically high magnetar formation rate (see \S\ref{sec:population}).

\subsection{Other scenarios}
Apart from the two most favorable scenarios discussed above, several other ideas have been proposed to explain these puzzling sources. \\

\noindent
$\bullet$ {\tt Isolated WD pulsar}. Rotating magnetic WDs emitting pulsed radio emission via magnetic dipole losses and pair production in the WD magnetosphere \citep{HurleyWalker2022, Katz2022}, as in a typical NS pulsar (first proposed by \citet{Zhang2005} for GCRT\,J1745$-$3009). This scenario naturally explains the slow rotation periods of LPTs and their large inferred population, but struggles to account for the bright and transient nature of the radio emission. Many nearby magnetic WDs are known, yet none show detectable pulsed radio emission. Furthermore, assuming a rotating dipole and magnetic fields of $10^8$--$10^9$\,G, the observed periods of LPTs are too long to generate the voltage gaps required for sustained pair production (see Figure~1 in \citet{2024ApJ...961..214R} and \S\ref{sec:radio_physics}). A hot subdwarf nature has also been proposed \citep{2022RNAAS...6...27L}, but faces similar difficulties and is excluded in several cases by IR and optical limits. \\

\noindent
$\bullet$ {\tt Binary neutron star systems}. Neutron stars with companions have been explored in several scenarios. Ordinary radio pulsars in binaries are unlikely to produce such bright radio pulses on such long timescales. However, \citet{Cary2025} proposed that a NS born in a close binary could acquire an accretion disk through interaction with the companion’s shock-inflated envelope before becoming unbound. Passage through the disk could temporarily alter its spin via a propeller phase. This model still requires a magnetar to power the radio emission and sustain an effective propeller (see \S\ref{sec:magnetar_scenario}). A related scenario was proposed by \citet{Mao2025}, involving a NS in a massive binary disrupted by a second supernova, transitioning from a high-mass X-ray binary phase before being ejected. In this case as well, the radio emission relies on a magnetar-like mechanism, and the implied formation rate would be even lower. \\

\noindent
$\bullet$ {\tt Primordial black holes}. \citet{PrimordialBH2024} proposed that radio emission could be produced by primordial black holes repeatedly passing through a host star, necessarily a WD in this framework. While intriguing, the authors find such encounters to be rare. \\

\noindent
$\bullet$ {\tt Self-lensed pulsar--black hole binaries}. \citet{Xiao2024} advanced the possibility that LPTs are pulsar--black hole binaries in which the pulsar signal is gravitationally self-lensed. This could generate apparently longer-period radio signals when the intrinsic pulsar is otherwise too faint to detect. However, the authors also find such events to be unlikely. Moreover, the observed polarization properties and microstructure of LPTs are difficult to reconcile with this scenario, as rapid intrinsic rotation should be detectable in high-time-resolution data. \\

\noindent
$\bullet$ {\tt Intermediate-mass black hole launching a jet}. \citet{Nathanail2025} investigated the possibility of an intermediate-mass black hole with a precessing accretion disk launching a Blandford--Znajek jet. While this model could potentially explain the periodicity and energetics, other characteristics such as the high linear polarization are difficult to reproduce. \\

\noindent
$\bullet$ {\tt Strange dwarf pulsars}. \citet{Zhou2025} proposed a compact object composed of a strange-quark-matter core surrounded by a normal-matter crust. This scenario could explain the long periodicities and, for magnetic fields of $10^6$--$10^{10}$\,G, produce radio emission via magnetospheric pair production, naturally accounting for the high polarization fractions. However, the transient radio activity and high-energy emission remain difficult to explain.

\section{Radio emission mechanisms}
\label{sec:radio_physics}

At the time of writing, the physical mechanism behind the bright radio pulses in LPTs has not yet been established. The brightness temperatures of LPTs are above the maximum allowed by incoherent emission, so the radio emission mechanism must be coherent, at least for the brightest pulses (see Figure\,\ref{fig:transient_plane}). 
However, several distinct physical ingredients are involved in producing such emission. In particular, it is useful to distinguish between the energy reservoir available to power the emission, the physical processes responsible for particle acceleration or for establishing a population inversion, and the coherent radiation mechanism that ultimately converts particle energy into radio emission. These components need not coincide uniquely in a given source, and similar observational properties may arise from different combinations of them. In the following, we discuss each aspect separately.

\subsection{Energy reservoirs powering LPT radio emission}

The first requirement for any viable model of LPTs is an energy reservoir capable of sustaining the observed radio luminosities. 
In most LPTs, the inferred radio luminosity exceeds the available rotational energy loss rate by several orders of magnitude in the isolated NS interpretations, and by a smaller but still significant factor in isolated white-dwarf interpretations \citep{HurleyWalker2022, 2024ApJ...961..214R}. This rules out purely rotation-powered emission analogous to that of ordinary radio pulsars as the main reservoir of all LPTs.

Alternative energy reservoirs have therefore been proposed. In the magnetar scenarios, magnetic energy stored in strong and possibly twisted magnetic fields may power particle acceleration and emission, either through gradual dissipation or episodic reconnection events (see i.e. \citet{Yang2025}). In WD systems, magnetic energy may also play a role, although the total available reservoir is typically smaller. In WD binary systems, an additional channel is provided by orbital and interaction energy, for example through unipolar induction or magnetospheric interaction between a WD and a companion star or its wind \citep{2016ApJ...831L..10G,2025ApJ...981...34Q, Yang2025arXiv250909224Y}.

At this level, the discussion concerns only the source of energy, and thus far what can be certainly excluded is only rotational power in a magnetized isolated compact object.

\subsection{Particle acceleration and population inversion}

The extremely high brightness temperatures observed in LPTs (Figure~\ref{fig:transient_plane}) imply a coherent radio emission process, which in turn requires particle distributions far from thermal equilibrium. Particle acceleration mechanisms must therefore operate to produce strongly anisotropic or population-inverted velocity-space distributions. Several physical processes may establish such conditions in LPTs, depending on the nature of the system.

In isolated compact objects, particle acceleration may occur in magnetospheric gaps or through magnetic reconnection; in NSs, this may further lead to pair cascades initiated by curvature or inverse-Compton photons. In magnetar-like environments, strong magnetic fields and dynamic magnetospheres can sustain repeated acceleration episodes and enhanced pair production. On the other hand, in WD binary systems, additional acceleration channels may arise at interaction sites between magnetospheres, winds, or currents induced by relative motion in a magnetic field.

Importantly, these processes are not radiation mechanisms themselves, but rather physical routes to creating the conditions necessary for coherent emission. For example, several types of anisotropic electron distributions such as loss-cone distributions, shell or horseshoe distributions, or distributions produced by current-driven instabilities, may generate a population inversion capable of driving maser-type coherent emission (e.g. electron cyclotron maser emission, ECME).

\subsection{Coherent radio emission processes}

Once a suitable particle acceleration mechanism is established and a population inversion or strongly anisotropic particle distribution is formed, several coherent radiation mechanisms may in principle operate.

One widely discussed possibility in the context of WD pulsars is ECME \citep{2025ApJ...981...34Q, Yang2025arXiv250909224Y}, which can naturally produce high brightness temperatures and strong polarization. ECME requires that the local plasma frequency be lower than the electron cyclotron frequency and is therefore sensitive to plasma density and magnetic field strength. These conditions may plausibly be met in low-density magnetospheric regions or in interaction zones in binary systems, although the escape of radiation from dense environments remains a challenge in some scenarios.

Relativistic versions of ECME have been explicitly proposed in the context of interacting WD binaries as a viable explanation for at least a subset of LPTs. In the unipolar-inductor framework developed by \citet{2025ApJ...981...34Q}, magnetic interaction between a rapidly rotating, magnetized WD and a low-mass companion can drive a particle flow powering the radio emission confined in a magnetic loop connecting the two stars. As a result, coherent radio emission can only be detected when the magnetic loop plane aligns with the line of sight, i.e. at conjunction, as is the case of ILT\,J1101$+$5521 for example. In that model, relativistic ECME is responsible for the radio emission, requiring a strongly magnetized WD and a mildly magnetized M dwarf. In this picture, the observed long radio periodicities may correspond to the spin, orbital, or beat periods of the system. A similar situation could be the case for GLEAM-X\,J0704$-$36, though in this case the geometry is different and the radio emission is detected at the ascending node of the orbit \citep{Rodriguez2025}. A recent work has successfully reproduced the above scenario via Particle-In-Cell simulations \citep{2025arXiv250909057Z}. 
More recently, \citep{Yang2025arXiv250909224Y} presented a related model in which LPTs originate from detached magnetic WD plus M-dwarf binaries in a pre--mCV evolutionary phase. In this scenario, asynchronism between the WD spin and the orbital motion sustains unipolar-induction or magnetospheric-interaction currents, while the low plasma densities expected at very low accretion rates allow ECME to operate efficiently. This work explicitly considered the loss-cone-driven maser (LCDM) as a particular form of ECME. This mechanism can efficiently generate coherent radio bursts with only sub-relativistic electrons, in contrast to the relativistic ECME discussed in \cite{2025ApJ...981...34Q}.
Both the unipolar inductor model \citep{2025ApJ...981...34Q} model and the magnetospheric interaction model \citep{Yang2025arXiv250909224Y} naturally produce a geometrically-beamed radiation pattern and a small duty cycle. The precise beaming angle in both scenarios is governed by the large-scale magnetic field geometry of the WD.

Another coherent emission candidate is coherent curvature radiation from bunched charges, analogous to mechanisms commonly invoked for radio pulsars and possibly magnetars. In this case, coherent emission arises from relativistic charges moving along curved magnetic field lines, provided that charge bunching can be sustained over appropriate spatial scales. While this framework can naturally reproduce pulsar-like polarization behaviour and microstructure, maintaining long-lived coherent bunches and, more generally, sustaining the required magnetospheric plasma supply, at the long characteristic periods of LPTs remains theoretically challenging \citep{1975ApJ...196...51R,1993ApJ...402..264C,2000ApJ...531L.135Z,Suvorov2023}.

In pulsar-like models, this difficulty is often discussed in terms of the ``death valley'' in the $P$--$\dot{P}$ diagram: not a sharp cutoff, but a region where the interplay between magnetic-field strength and geometry, particle acceleration, and radiative processes may suppress efficient pair production and hence quench coherent radio emission. While such boundaries have historically been used to explain the scarcity of very long-period radio pulsars, LPTs challenge this paradigm if interpreted as pulsar-like sources (see Figure~\ref{fig:p_pdot}). A similar constraint arises if the sources are instead interpreted as isolated white-dwarf systems (see Figure~1 in \citealt{2024ApJ...961..214R}).

In the magnetar scenario, an interesting investigation by \citet{2024MNRAS.533.2133C} explored low-altitude pair production beyond curvature-radiation death lines. In this framework, plastic deformation of the NS crust or thermoelectric effects associated with strong temperature gradients in slowly rotating, long-period magnetars can induce mild twists in the external magnetic field. Such twisted magnetospheres may support parallel electric fields and voltage gaps, which can accelerate particles and trigger pair cascades via resonant inverse-Compton scattering (RICS) or curvature radiation, potentially powering coherent radio emission. These twist-driven cascades require magnetar-strength magnetic fields and long spin periods in order to reproduce the observed luminosities, timescales, and timing properties.

The presence of potential inter-pulses in ASKAP\,J1839$-$0756 \citep{2025NatAs...9..393L}, state changes in ASKAP\,J1935$+$2148 \citep{2022NatAs...6..828C} and the linear-to-circular polarization conversion and drifting sub-pulses in GPM\,J1839$-$10 \citep{Men025} have strongly argued for a similarity to pulsars and magnetars. However, the recent detection of an orbital period in this latter source and its successful modeling with a binary WD/M-dwarf system \citep{2025arXiv250906315H} casts doubt on the univocal relation of these observed properties with a magnetized NS magnetosphere, and opens the possibility of being produced by the interaction with the magnetized wind of a low-mass star.

Other coherent processes, such as plasma emission or synchrotron maser emission under relativistic conditions, have also been discussed, though their applicability to the full range of LPT phenomenology is still under investigation. 

Regardless of the specific coherent emission process, the strong polarization and apparent beaming observed in LPT radio pulses provide important constraints that must be interpreted with care. Geometrical beaming is expected whenever relativistic particle streams are confined to ordered magnetic field lines and therefore arises primarily from particle kinematics and magnetic geometry, rather than being an intrinsic property of any particular radiation mechanism. Consequently, neither ECME nor curvature radiation nor unipolar-induction models uniquely produce geometrically beamed emission. Instead, both isolated magnetospheric models and binary interaction scenarios may give rise to strongly beamed radiation, provided that relativistic particle flows develop along structured magnetic fields. Differences in polarization fraction, position-angle behaviour, and pulse morphology may then reflect variations in geometry, viewing angle, and plasma conditions rather than fundamentally different emission mechanisms.

\section{Population constraints}
\label{sec:population}

As the field develops, it may become possible to analyse the distribution of the properties of the entire LPT population in order to better determine their progenitors (see e.g. \cite{2024ApJ...968...16G} and \cite{2022AJ....163...69B} for a pulsar population study and an observationally driven FRB population analysis). However, at the time of writing, the known population of LPTs remains small and is still drawn from highly heterogeneous surveys and observing strategies.

Following the earliest discoveries, the distribution of the initial class members in the Galactic plane was initially suggestive of a young population of compact objects. \cite{Dobie2024} noted that the LPTs were confined to the thin disc, with a distribution dissimilar to the wide range of Galactic latitudes over which WDs \citep{2021MNRAS.508.3877G} and old pulsars \citep{2005MNRAS.360..974H} are found, but instead more closely resembling the very low latitudes at which canonical magnetars are observed \citep{2014ApJS..212....6O}. However, further findings have not confirmed this preliminary suggestions, and given that approximately 85\% of the stars in the Milky Way reside within the thin disc, it is perhaps not surprising that the eight known LPTs at that time were located there. Since then, sources such as CHIME/ILT\,J1634$+44$ have been discovered (see also Figure\,\ref{fig:positions}) at heights of 0.7--3\,kpc above the Galactic plane, which are difficult to reconcile with a thin-disc-only population \citep[scale height $279$\,pc;][]{2023Galax..11...77V}. This highlights the necessity of a significantly larger sample before robust population-based inferences can be drawn.

Fortunately, larger samples are likely to become available over the coming decade. The original discovery of GCRT\,J1743$-$3009 by \cite{2005Natur.434...50H} was followed by a nearly two-decade interval during which no similar sources were identified. The wide fields of view of SKA precursor instruments, coupled with steadily improving capabilities over the past ten years, have since enabled the discovery of a much larger population. That many of these telescopes have now been operating for over a decade \citep[e.g. MWA and LOFAR][]{2013A&A...556A...2V,2013PASA...30....7T}, and that some LPTs have been detected in historical data \citep[e.g. GPM\,J1839$-10$ in archival VLA and GMRT observations][]{2023Natur.619..487H}, yields two important insights: (i) archival observations from radio facilities are likely to contain additional, previously unidentified LPTs; and (ii) determining the timing and duration of activity windows for newly discovered sources will be considerably facilitated by the rapidly growing volume of archival data.

In this sense, current surveys have likely uncovered only the tip of the iceberg of this source population. It is presently unknown whether a substantial population of faint or only briefly active LPTs exists.
While anecdotal, such events illustrate the observational incompleteness of current surveys. These sources are intrinsically more difficult to detect, and the existing literature is naturally dominated by single-object discoveries. Constraining (or placing stringent limits on) such populations will be essential for determining key physical properties. For example, if the luminosity function exhibits a low-luminosity cut-off, this may imply a minimum energy threshold at which the as-yet unknown radio emission mechanism can operate. Conversely, a large population of briefly active sources could indicate characteristic timescales for magnetic-field decay or transformation in magnetar-based scenarios \citep{2024MNRAS.533.2133C}, or for charged-particle outflows in WD--M-dwarf systems \citep{2025ApJ...981...34Q,Yang2025arXiv250909224Y}. The repeated transient nature of LPTs therefore motivates repeated monitoring of as wide an area as possible, whereas for continuously active sources, periodic integration of a given field to the required sensitivity would suffice for detection.

If a low-luminosity LPT population exists, then current single-pulse radio searches \citep{2023MNRAS.523.5661W,2024MNRAS.531.4805D,2025arXiv250906315H} will ultimately prove insufficient. In such cases, folding-based searches would be required, although these currently pose significant technical challenges on the long timescales relevant for LPTs. Where an optical counterpart is known, folding searches could initially be guided by optical parameters; however, such approaches would not constitute blind searches and would inherit selection biases from the optical surveys themselves. Folding-based methods would therefore probe a complementary region of parameter space rather than replacing single-pulse techniques.

Optimizing survey strategies both for discovering LPTs and for conducting robust population studies remains challenging, as the duration and duty cycle of their radio activity are poorly constrained, and eventually variable even on a single source. Current surveys are strongly biased towards specific timescales, periodicities, and flux ranges due to their observing strategies, greatly complicating attempts to infer the properties of the underlying population. Any future population analysis will therefore require explicit modelling of survey selection functions, including cadence, sensitivity, frequency coverage, and sky exposure.

One conclusion is nevertheless clear: the discovery rate of LPTs has increased rapidly over the past $\sim$3 years, reaching rates of $\gtrsim$4--5 sources per year. While this trend likely reflects improved survey coverage and sensitivity rather than an intrinsic increase in source formation, it nonetheless suggests that LPTs may represent a relatively common Galactic population, potentially analogous in prevalence to WD--M-dwarf binary systems. Given their reduced radio activity windows (see Figure\,\ref{tab:lpt_radio}) and the substantial observational biases discussed above, the true Galactic population may be significantly larger than currently observed.

Ultimately, search methodologies will continue to evolve as the field matures, driven by both the excitement of exploring this new region of parameter space and the need to adapt to the strengths and limitations of different radio facilities. Deconvolving the resulting selection biases in order to recover the underlying astrophysical distributions will be a challenging and rewarding endeavor in the coming years.

\section{Summary and Future Prospects}

Since their initial discovery in 2022, enabled by the introduction of new methods to identify periodic radio sources in imaging data, LPTs have been steadily uncovered in both archival and ongoing radio surveys. Several of these objects are now thought to be associated with WD binary systems containing low-mass companions, with proposed links to detached WD pulsar systems such as AR~Sco--like sources, and potentially to CVs. The physical mechanism responsible for the bright, periodic radio bursts remains uncertain, although it is likely related to interactions between the wind of the low-mass companion and the magnetic field of the WD. Alternative explanations, including a magnetar origin in which the spin period has been significantly lengthened through supernova fallback accretion, may still apply to a minority of sources. In addition, recent discoveries of variable X-ray activity from LPTs, reminiscent of CV or magnetar outbursts, further highlight the diversity of their phenomenology. Continued multiwavelength follow-up observations and the discovery of additional LPTs will therefore be essential for resolving these possibilities and for achieving a comprehensive understanding of the physical origin of their bright, periodic radio emission. Predictions for GW emission have been studied, adding these sources as potential targets for space-based GW interferometers \cite{2025ApJ...991..134S}.

Taken together, the considerations discussed above suggest that LPTs may plausibly be powered by multiple emission pathways. In magnetar-based interpretations, magnetic energy provides the primary energy reservoir, particle acceleration may be triggered by magnetospheric activity or magnetic reconnection, and coherent radio emission could arise through curvature radiation or relativistic ECME processes. In WD binary systems, orbital or interaction energy may instead dominate, with particle acceleration occurring in magnetospheric interaction regions and coherent emission most likely produced via maser or plasma processes.

At present, the available observational constraints do not uniquely select a single combination of energy reservoir, particle acceleration mechanism, and coherent emission process applicable to the entire LPT population. It is therefore plausible that they have a single unifying emission scenario, but more than one physical pathway remains possible. Nevertheless, current data already disfavour emission mechanisms that rely on standard pulsar-like scenarios that would require a sustained acceleration gap powered by rotational energy. Such models are difficult to reconcile with the long spin periods and limited rotational energy budgets of LPTs, in any plausible astrophysical scenario.

\bigskip
\noindent
{\bf Acknowledgements}
N.R. is supported by the European Research Council (ERC) via the Consolidator Grant “MAGNESIA” (No. 817661) and the Proof of Concept ``DeepSpacePulse" (No. 101189496), by the Catalan grant SGR2021-01269, the Spanish grant PID2023-153099NA-I00, and by the program Unidad de Excelencia Maria de Maeztu CEX2020-001058-M. N.H.-W. is the recipient of an Australian Research Council Future Fellowship (project number FT190100231). M.C. acknowledges support of an Australian Research Council Discovery Early Career
Research Award (project number DE220100819) funded by the Australian Government. 

We thank Fengqiu Adam Dong, Ingrid Pelisoli, Matteo Imbrogno, Francesco Coti Zelati, Noel Castro-Segura, and Yuan-Pei Yang for useful comments on the manuscript, as well as Bing Zhang, Simone Scaringi, Thomas Tauris, Christian Knigge, Sam McSweeney, Csanad Horvath, Ziteng (Andy) Wang, Kaya Mori, Andrea Possenti, Marta Burgay and Domitilla DeMartino for interesting discussions on LPTs. NR thanks Benoit Cerutti, Christo Venter, and all the participants and organizers of the Les Houches Workshop 2025 "Feeling the pull and the pulse of relativistic magnetospheres" for insightful discussions that helped shaping this manuscript. We acknowledge the very useful suggestions of the referee that largely improved this work.

\bigskip
\noindent
{\bf Data sharing}
Data and jupyter notebooks for reproducing plots and tables are collected here: \url{https://github.com/nandarea/LPT_Review_2026.git} .

\bibliographystyle{elsarticle-harv} 
\bibliography{refs_keydedup_fixed.bib}

\end{document}